Noble gases in high-pressure silicate liquids: A computer simulation study.

Bertrand Guillot and Nicolas Sator

*Laboratoire de Physique Théorique de la Matière Condensée, Université Pierre et Marie Curie (Paris 6),*

*UMR CNRS 7600, case courrier 121, 4 place jussieu, 75252 Paris cedex 05, France.*

**ABSTRACT**

The test particle method has been used in conjunction with molecular dynamics simulations to evaluate the solubility of noble gases in silicate melts of various composition. At low pressure the calculated solubility constants (the inverse of the Henry's constant) are in excellent agreement with data of the literature. In particular it is found that the solubility constant (i) decreases when the size of the noble gas increases, (ii) decreases from silica-rich to silica-poor composition of the melt, and (iii) is positively correlated with the temperature. Moreover it is shown that the solubility is governed primarily by the entropic cost of cavity formation for inserting the noble gas into the melt and secondarily by its solvation energy. Interestingly, the behaviour of these two contributions differ from each other as the entropic cost of cavity formation increases strongly with the size of the solute atom to insert whereas large atoms are better solvated than small ones. Considerations of thermodynamics show that the weight fraction of a noble gas in a silicate melt coexisting with its parent fluid at T and P is equal to $n_g \gamma_m / n_m \gamma_g$ , where $n_g$ and $n_m$ are the densities of the two coexisting phases (gas and melt, respectively) and where the solubility parameters $\gamma_m$ and $\gamma_g$ express the probability of inserting the noble gas atom in the melt and in the parent fluid, respectively. The $\gamma_m$ and $\gamma_g$ decrease drastically when the pressure is increased and the noble gas solubility at high pressure is the result of a balance between these two quantities. Here again, the pressure behaviour of $\gamma_m$ and $\gamma_g$ is dominated by the pressure dependence of the entropic cost of cavity formation, the energetic contribution being of minor importance but not negligible at high pressures. With all melt composition investigated here (silica, rhyolite, MORB and olivine), the calculated solubility curves exhibit the same qualitative behaviour with pressure; a steep rise culminating in a broad maximum followed by a gradual decrease of the solubility at higher pressure. At variance with LHDAC experiments (Chamorro et al. (1996, 1998) and Bouhifd et al. (2006, 2008)) where a Ar solubility drop is observed at about 50 kbar in silica and molten olivine and in the pressure range ~100-160 kbar with other melt composition, we do not find such a sudden change of the solubility.

# 1. Introduction

Noble gases and their isotopes (primordial or radiogenic) are important tracers of the Earth's mantle dynamics (Allègre et al., 1983; Graham, 2002; Albarède, 2008; Gonnermann and Mukhopadhyay, 2009; Coltice et al., 2011) and of the formation of terrestrial planets and their atmosphere (Harper and Jacobsen, 1996; Pepin and Porcelli, 2002). The details of the noble gas transport from deep mantle reservoirs to the exosphere are still debated (Gonnermann and Mukhopadhyay, 2009). A key quantity necessary for evaluating the efficiency of degassing scenarios is the solubility of noble gases in the silicate melt and particularly its evolution with depth, i.e. with the pressure. For example, noble gas partition coefficients between mineral and melt depend on the solubility ratio in these two phases whereas the partitioning of noble gases between the melt and the $CO_2$ bubbles depends on the noble gas solubility in the melt as compared with that in the supercritical $CO_2$ phase contained in the bubbles. If bubble nucleation occurs at shallow depth (e.g. very near the seafloor where P~300-400 bar) the noble gases present in the melt will partition preferentially into the bubbles because of the lower solubility of noble gases in the silicate melt (Jambon et al., 1986). In contrast, if bubble nucleation starts at greater depth, the high density of the $CO_2$ fluid may render the transfer of noble gases from the melt to the vesicles more difficult (Sarda and Guillot, 2005). In the same way, if partial melting occurs in the upper mantle (~80-300 km after Dasgupta and Hirschmann, 2006), the solubility of noble gases in the resulting melt may be so low that it can balance the solubility in the crystal. For example, recently the commonly assumed large incompatibility of noble gases in the mantle has been challenged (Watson et al., 2007) but these results have been contested since then (Cassata et al., 2011).

It is, therefore, important to know the solubility of noble gases in silicate liquids with pressure and with melt composition. Melt composition dependence of noble gas solubility is well documented at low pressures up to the kbar range (Kirsten, 1968; Hayatsu and Waboso, 1985; Jambon et al., 1986; Lux, 1987; Broadhurst et al., 1990,1992; Roselieb et al., 1992; Carroll and Stolper, 1993; Shibata et al., 1998; Walter et al., 2000; Mesko and Shelby, 2002; Miyazaki et al., 2004; Marrocchi and Toplis, 2005; Tournour and Shelby, 2008a,b; Iacono-Marziano et al., 2010). Higher-pressure data are more limited (White et al., 1989; Montana et al., 1993; Chamorro et al., 1996, 1998; Schmidt and Keppler, 2002; Bouhifd and Jephcoat, 2006; Bouhifd et al., 2008). In the latter studies the noble gas solubility (in weight percent) increases almost linearly with the pressure in the 0~30 kbar range, before to level off at a higher pressure. However, in some melts (e.g. silica and molten olivine) the argon solubility drops abruptly above ~50 kbar (Chamorro et al., 1996, 1998; Bouhifd and Jephcoat, 2006; Bouhifd et al., 2008) whereas in other melt compositions (e.g. anorthite,, sanidine, haplogranite and haplotholeite) the argon solubility reaches a plateau value (Chamorro et al., 1996; Schmidt and Keppler, 2002) before an abrupt drop between 100 and 160 kbar depending on melt composition (Bouhifd and Jephcoat, 2006; Bouhifd et al., 2008). If correct, this solubility drop may have important geochemical implications as it suggests that below some depth (i.e. above a threshold pressure) partial melting could not be an efficient way for noble gas extraction from the solid mantle.

From a theoretical standpoint these conclusions are problematic because from statistical physics using a hard sphere model often used in liquid state physics, it has been shown (Sarda and Guillot, 2005; Guillot and Sarda, 2006) that the concurrent compaction of coexisting fluid and melt is responsible of the quasilinear (Henry-like behaviour) increase of the noble gas solubility with pressure up to ~30 kbar. With further pressure increase the solubility levels off in the 40~80 kbar range before to decrease gradually (and not abruptly) at higher pressure. It has been argued for Al-bearing silicate melts (see Bouhifd et al., 2008) that the pressure onset of the Ar solubility drop correlates with the Al/(Al+Si) ratio, with the possibility that a coordination change of Al (from 4-fold to 5- and 6-fold coordination) in the corresponding pressure range (80~160 kbar) could be the origin of the solubility drop. However this explanation is irrelevant for Al-free melts like silica and molten olivine which both exhibit an Ar solubility drop at nearly the same pressure (~50 kbar). This finding could be coincidental because the two melts are very different, silica being fully polymerized and characterized by high Ar contents before the solubility drop whereas molten olivine is a very depolymerized melt and can accomodate only very low Ar contents (the solubility maximum in silica is ~25 times higher than in molten olivine). Chamorro et al. (1998) have suggested that the hole size distribution in these two melts shrinks so much under pressure that above ~50 kbar their structure cannot accommodate a solute particle with the size of Ar. Although this explanation is tempting, it is based upon a very crude approximation of the melt structure where the free volume accessible to the solute particles is given by the difference between the specific volume and the volume occupied by the elements of the melt. This model is useful at low pressures but it cannot take into account all the complexity of the densification of the melt at high pressures such as coordination change of network former ions and the evolution of the T-O-T bond angle distribution (with T=Si, Al). Moreover it ignores the densification of the fluid phase in contact with the melt, a mechanism that is essential because the solubility is a function of the variation of the noble gas chemical potential. In fact a more rigorous approach to describe the incorporation of noble gases in silicate melts can be obtained from statistical theory (Sarda and Guillot, 2005; Guillot and Sarda, 2006) and from molecular simulations (Guillot and Guissani, 1996b; Zhang et al., 2010).

For instance, Guillot and Guissani (1996b) evaluated the free energy of insertion of noble gases in fused silica by molecular dynamics (MD) simulation using the test particle method (TPM) originally proposed by Widom (1963). From these calculations the incorporation of a noble gas into the melt is governed principally by the entropy of cavity formation, the solvation energy of the solute atom being of secondary importance. A direct consequence of this result is that small atoms (e.g. He and Ne) are preferentially solvated with respect to large ones. More recently, Zhang et al. (2010) evaluated by MD simulation the distribution of interstitial voids in the structure of fused silica in modeling the oxygen and silicon atoms as exclusion spheres of given radii. The Ar solubility is calculated from a statistical model depending on the hole size distribution and the fugacity of the coexisting noble gas fluid. The pressure-dependent Ar solubility in fused silica calculated in this way is in a qualitative agreement with the experimental data of Chamorro et al. (1996) and Bouhifd et al. (2008). Although these results are quite interesting, they have to be viewed with some caution because neither the softness of the atoms in response to densification nor the solvation energy and its evolution with pressure are taken into account. So it is convenient to reinvestigate by computer simulation the solubility of Ar in fused silica and to extend the calculation to other noble gases and silicate compositions to shed light on the mechanisms of incorporation of noble gases in silicate melts under pressure.

In the present study we have evaluated by MD simulation the solubility of noble gases in silicate liquids from acidic to ultrabasic compositions (silica, rhyolite, mid-ocean ridge basalt, olivine, and enstatite). The objective is to have an accurate theoretical tool for predicting the evolution of the noble gas solubility with melt composition, pressure and nature of the noble gas. The crux of the calculation is the implementation of the TPM (test particle method) in a MD simulation scheme to calculate the chemical potential of the noble gases in the two phases (silicate melt and rare gas fluid) assumed to coexist with each other at given (P,T) conditions.

## 2. Method of calculation

### 2.1 Solubility and test particle method

The equality of the chemical potentials of a rare gas atom in coexisting melt and fluid at given (P,T) conditions, leads to the following relationship,

$$\rho_m^0/\rho_g = e^{-(\mu_m^{ex} - \mu_g^{ex})/k_B T} = \gamma_m / \gamma_g \qquad (1)$$

where $k_B$ is the Boltzmann constant, $T$ the temperature, $\rho_m^0$ and $\rho_g$ are the number densities (number of atoms per unit volume) of the noble gas in the silicate melt and in the fluid phase, respectively, and where $\mu_m^{ex}$ and $\mu_g^{ex}$ are the excess chemical potentials of the solute in the corresponding phases (notice that the ideal parts of the chemical potentials cancel out because they are identical in the two phases). The quantity $\gamma_i = e^{-\mu_i^{ex}/k_B T}$ (with $i = m$ or $g$) is named the solubility parameter in the corresponding phase. Usually the quantity actually measured in a real experiment is the weight fraction $X_W$ of noble gas in the melt. In using Eq.1 the weight fraction of noble gas can be written as,

$$X_W = L_W/(1+L_W) \qquad (2)$$

where $L_W = n_g \, \gamma_m \, / \, n_m \, \gamma_g$ , and $n_m$ and $n_g$ are the densities of the silicate melt and of the fluid phase at the condition of interest. Notice that generally $L_W \ll 1$ and, therefore, $X_W \sim L_W$ . If one prefers to deal with mole fraction, $X$, of noble gas in the melt then Eq.2 becomes,

$X = L /(1+L)$             (3)

where $L = \rho_g \, \gamma_m \, / \, \rho_m \, \gamma_g$ , and $\rho_m$ and $\rho_g$ are the number densities of the silicate melt and of the fluid phase, respectively. The above expressions are valid at any pressure and temperature. It is only at very low pressure that the fluid phase can be considered as nearly ideal. In that case $\gamma_g \sim 1$ and $P_g \sim \rho_g k_B T$ , and the mole fraction of noble gas in the melt can be approximated by the well known Henry's law,

$X = P_g \, S$             (4)

where $S = \gamma_m \, / \rho_m k_B T$ is the solubility constant (the inverse of the Henry constant). In practice the solubility constant of noble gases in silicate melts is in the range $10^{-5}$-$10^{-8} \, bar^{-1}$ (for a data compilation see Paonita et al. (2005)). At higher pressure, in the kbar range and above, neither the pressure dependence of $\gamma_g$ and $\gamma_m$ resulting from the compression of the coexisting phases, nor the actual variation of the densities of the two phases can be neglected in Eqs.(2) and (3).

In statistical mechanics the excess chemical potential of a solute particle in a solvent can be written under a form originally proposed by Widom (1963) (for a textbook see Frenkel and Smit, 2002),

$\mu^{ex} = - \, k_B T \, ln < e^{-\psi/k_B T} >_N$             (5)

where $\psi = (U_{N+1} - U_N)$ is the potential energy difference between a mixture composed of $N$ solvent particles plus the solute and the pure solvent ($N$ particles). In the case where the potential energy of the system is pairwise additive, $\psi$ is nothing but the solute-solvent interaction energy. A remarkable consequence of Eq.5 is that the canonical average $< \cdots >_N$ is taken over the configurations of the pure solvent, the solute particle acting as a ghost (or a test ) particle. In practice the solute particle is inserted at random in the solvent configurations generated by MD simulation and the Boltzmann factor $e^{-\psi/k_B T}$ is averaged over all events. However, at liquid-like densities (e.g. in a silicate melt or in a compressed rare gas fluid) the inserted test particle has a very high probability to be in close contact with solvent particles. Therefore, the solute-solvent interaction energy is strongly repulsive ($\psi$/

$k_B T \gg 1$). The result is a vanishingly small contribution of these events to the average ( $e^{-\psi/k_B T} \ll 1$ ). What is needed for practical use of this method is a numerical recipe for detecting quickly undesirable positions of the test particle and for locating cavities that can accommodate the solute (here a noble gas atom), cavities that appear and disappear at the mercy of the solvent fluctuations.

Such a recipe has been developed by Deitrick et al. (1989) and makes the test particle method very effective to evaluate the chemical potential of a solute in a liquid (e.g. Paschek, 2004; Shah and Maginn, 2005). The procedure consists in dividing the simulation box containing the solvent particles into small cubelets. A cubelet was marked as occupied if its center was located within the highly repulsive region of any solvent atoms. For each noble gas - ion pair (X-i) in the silicate melt this corresponds to a cut off radius $r_{X-i}^{cut}$, defined in such a way that if a test particle was located at the center of an occupied cubelet the value of the corresponding Boltzmann factor $e^{-\psi/k_B T}$ would contribute virtually nothing to the total average (in practice an upper bound of the cutoff radius $r_{X-i}^{cut}$ can be estimated by recording the shortest interatomic distance $d_{X-i}^{min}$ reached by a pair (X-i) during a MD simulation run where a noble gas atom is diluted in a silicate melt). For a MD-generated atomic configuration of the solvent, a map of occupied cubelets is evaluated (defining the excluded volume) and the test particle is inserted only into the unoccupied cubelets (the free volume). Hence the average of the MD configuration is given by,

$$\overline{e^{-\psi/k_B T}} = \frac{1}{N} \sum_{i=1}^{N_u} e^{-\psi_i/k_B T} \qquad (6)$$

where $N$ is the total number of cubelets (in general $41^3$ or $161^3$), $N_u$ the number of unoccupied cubelets and where the index $i$ runs over all unoccupied cubelets, $\psi_i$ being the interaction energy between the test particle in the cubelet i and all the atoms of the solvent.

The excess chemical potential (see Eq.5) is evaluated by averaging $\overline{e^{-\psi/k_B T}}$ over thousands of MD steps (see next section for computational details). When the solvent under investigation is very dense (e.g. a silicate melt at very high pressure) and/or the solute atom is large (Xe for instance) the proportion of unoccupied cubelets can be as small as $10^{-5}$ or less and the statistics becoming poor the result is unreliable. This problem of statistical inaccuracies occurred with Xe in silicate melts for pressure above ~50 kbar, and with Ar above 150 kbar, wheras the results with He and Ne are quite reliable up to 200 kbar. A more detailed discussion on the statistics is given in Appendix A.

Another advantage of the test particle method (TPM) is the possibility of evaluating the average solute-solvent energy ($E_\mu$) and the entropic cost for inserting the solute into the solvent ($\Delta S_\mu$). These two quantities are related to the excess chemical potential through the following relationship,

$$\mu^{ex} = E_\mu - T\Delta S_\mu \tag{7}$$

where $\mu^{ex}$ is obtained from Eq.5 and where $E_\mu$ and $T\Delta S_\mu$ are given by,

$$E_\mu = <\psi e^{-\psi/k_B T}>_N / <e^{-\psi/k_B T}>_N \tag{8}$$

$$T\Delta S_\mu = k_B T \, ln <e^{-\psi/k_B T}>_N + <\psi e^{-\psi/k_B T}>_N / <e^{-\psi/k_B T}>_N \tag{9}$$

Note that, from a thermodynamic standpoint, the above relationships are exact (for details see Yu and Karplus (1988)) and are very effective to evaluate $E_\mu$ and $\Delta S_\mu$ by the TPM. Thus the absolute magnitude of the excess chemical potential expresses an energy-entropy compensation. For example, in molecular liquids the solvation energy $E_\mu$ of apolar species (e.g. noble gases in water, see Guillot and Guissani (1993) and Graziano (1999)) can be negative or positive according to the temperature, the density of the solvent and the nature of the solute. In contrast the entropic contribution $\Delta S_\mu$ is generally negative and tends to inhibit the incorporation of these species in the solvent. For the following it is useful to express the solubility parameter $\gamma_i$ (where $i = m$ or $g$) in terms of an energetic ($\gamma_i^E$) and an entropic ($\gamma_i^S$) contribution namely,

$$\gamma_i = \gamma_i^E \times \gamma_i^S \tag{10}$$

where $\gamma_i^E = e^{-E_\mu^i/k_B T}$ and $\gamma_i^S = e^{\Delta S_\mu^i/k_B}$.

## 2.2 Computational details

To evaluate the noble gas solubility by the TPM one needs to generate by MD simulation a number of atomic configurations which are representative of the silicate melt and of the rare gas assumed to coexist at a given state point. The success of a MD simulation relies on the accuracy of the force field used to describe the interactions between atoms or ions in the system under investigation. In the present case there are three different sets of interaction involved: noble gas-noble gas interaction, silicate-silicate interaction and noble gas-silicate interaction. The derivation of the corresponding interaction parameters is described in Appendix B. The potential parameters are listed in Table 1 for noble gas-noble gas interactions and in Table 2 for noble gas-silicate interactions.

All molecular dynamics simulations were performed with the DL_POLY 2.0 code (Smith and Forrester, 1996). The equations of motions for ions (in the case of silicate melts) and noble gas atoms (in the case of the supercritical rare gases) were solved with a time step of 1 fs ($10^{-15}$ s) by the Verlet's algorithm. The simulation box is cubic with periodic boundary conditions and contains 500 atoms when simulating rare gas fluids and 1,000 ions for silicate melts (see Table 1 in Guillot and Sator (2007a) for chemical compositions of rhyolite, MORB, and San Carlos olivine melts investigated here). The long range coulombic interactions between ions were accounted for by a Ewald sum with $\alpha L=5\sim7$ where $\alpha$ is the width of the charge distribution on each ion and $L$ the length of the simulation box ($L\sim20A$). For the two systems the calculations were first performed in the isothermal isobaric ensemble (N,P,T) for equilibration and next were carried on in the microcanonical ensemble (N,V,E) for generating long production runs (up to 100 ns or $10^8$ MD steps). The uncertainties on each derived state are $\Delta P/P \sim \pm 1\%$ and $\Delta T/T \sim \pm 2\%$.

To evaluate accurately the solubility of noble gases in silicate melts the length of the simulation runs have to be sufficient to ensure that the melt is fully relaxed. A rapid evaluation of the relaxation time using the Maxwell relation ($\tau_{relax} = \eta/G_\infty$ where $\eta$ is the melt viscosity and $G_\infty$ the shear modulus at infinite frequency, $\sim 10^{10}$ Pa, Dingwell and Webb (1989)) shows that a simulation run of 10 ns is required to relax a silica melt at 2600K (e.g. $\eta^{sim} \sim 26$ Pa.s and $\tau_{relax}^{sim} \sim 2.6$ ns) and a rhyolitic melt at 1673 K (e.g. $\eta^{sim} \sim 16$ Pa.s and $\tau_{relax}^{sim} \sim 1.6$ ns), whereas for basaltic and ultrabasic melts structural relaxation is achieved on a time scale much shorter than a nanosecond (e.g. $\eta^{sim} \sim 0.6$ Pa.s and $\tau_{relax}^{sim} \sim 0.05$ ns for MORB at 1673 K and $\eta^{sim} \sim 0.05$ Pa.s and $\tau_{relax}^{sim} \sim 0.005$ ns for molten olivine at 2273 K). Note that the simulated silica melt and rhyolitic melt are less viscous than the real substances at the same thermodynamic conditions, because of a failure of the force field used to describe these melts (see Carré et al., 2008; Guillot and Sator, 2007a). In contrast the viscosity of basic and ultrabasic melts are reproduced well by the simulations. However this issue is immaterial for the evaluation of the noble gas solubilities. For rare gas fluids, the structural relaxation time is not limiting because it is very short (~1ps) in the supercritical states investigated here. Nevertheless an accurate evaluation of the solubility parameters of noble gases in their own fluid by the TPM, and especially at high pressures, needs to sample a large number of atomic configurations (see Appendix A). Thus even for the rare gas fluids, production runs as long as ~10 ns have been performed.

As in a real experiment, the basic idea of the theoretical approach is to describe the coexistence between a rare gas fluid and a silicate melt. However, in the simulation procedure the two phases are simulated independently from each other at the same (P,T) conditions and the computer generated atomic configurations were stored every $k$ steps (typically $k = 1,000$ MD steps) to be sampled subsequently with the TPM. It is noteworthy that this approach is exact in the infinite dilution limit when noble gases are trace elements in the silicate melt (for more theoretical details see Frenkel and Smit (2002)). For higher noble gas concentrations, this approximation is valid as long as the presence of the noble gases in the coexisting silicate melt has a negligible effect on the thermodynamic properties of the melt, and also that the traces of silicate elements diffusing into the rare gas fluid have a negligible effect on the fluid properties. We will see later on that these assumptions can be validated by performing a simulation where the rare gas fluid is explicitly in contact with the silicate melt and in comparing the solubility of the noble gases so obtained with the results of the TPM.

# 3. Results and Discussion

## 3.1 Solubility constants

The solubility constants (see Eq.4) of He, Ne, Ar, Kr, Xe and Rn in rhyolite, MORB, olivine, and enstatite melt have been evaluated by the TPM in the temperature range 1673-2273K. The results are listed in the Table of Appendix C (supplementary data) and a detailed comparison with solubility data of the literature for He, Ne, Ar and Xe is presented in Fig.1 as function of temperature and melt composition (results for Kr and Rn are not presented in Fig.1 because solubility data for Kr are scarce or absent in the case of Rn). The general agreement between simulation results and solubility data is very satisfying. Albeit we have used the solubility data in tholeiitic basalt melts (Jambon et al., 1986, Lux, 1987) to adjust the L-J parameters associated with the noble gas-silicate interactions (see Appendix B), the agreement obtained for acidic (e.g. rhyolite) and ultrabasic (e.g. olivine and enstatite) melts is a stringent test of the degree of accuracy of the calculation. However, one has to emphasize that the experimental data are somewhat scattered because of slight compositional differences between related melts or differences in experimental set up and method of analysis (for instance a deviation as large as a factor of 2 is usual for Ar in basaltic melts, see the data compilation of Carroll and Stolper (1993) and Paonita (2005)). Furthermore the high diffusivity of He could lead to underestimate its solubility (gas loss when cooling the sample, see Jambon et al. (1986) and Shibata et al. (1998)) whereas the solubility of Xe could be overestimated by atmospheric contamination because of its high potential of adsorption (for a discussion see Jambon et al. (1986)).

The solubility of noble gases exhibits three main tendencies (Fig.1): (i) The heavier the noble gas the lower the solubility in a melt of given composition, (ii) The more depolymerized the melt, i.e. the higher the NBO/T ratio (nonbridging oxygen per tetrahedrally coordinated cation, ~0.016 in rhyolite, ~0.33 in MORB and ~4.0 in olivine), the smaller the noble gas solubility, and (iii) the higher the temperature the larger the solubility. With regard to point (i), the calculations indicate that the solubility constant, $S = \gamma_m / \rho_m k_B T$ with $\gamma_m = \gamma_m^S \times \gamma_m^E$ (see Eq.10), is governed mainly by the entropic cost of cavity formation ($\gamma_m^S = e^{\Delta S_\mu^m/k_B}$) to accommodate the noble gas atom in the melt and secondarily by the solvation energy ($\gamma_m^E = e^{-E_\mu^m/k_B T}$). This is illustrated in Fig.2 where the solubility parameter in the melt, $\gamma_m$, its entropic contribution, $\gamma_m^S$, and its energetic contribution, $\gamma_m^E$, are shown as function of the van der Waals diameter of the noble gas. Clearly the behaviour of $\gamma_m$ is driven by $\gamma_m^S$ which decreases when the size of the noble gas increases, whereas the role of $\gamma_m^E$ is minor although it increases with the size of the noble gas (notice that $\Delta S_\mu^m$ is always negative whereas $E_\mu^m$ is positive for He and Ne, and negative for Ar, Kr, Xe and Rn). In the same way the variation of the solubility constant with melt composition is dominated by that of $\gamma_m^S$. Thus the higher the degree of depolymerization the denser the melt and the larger the entropic cost to accommodate a noble gas atom of a given size ($\gamma_{rhyolite}^m > \gamma_{MORB}^m > \gamma_{olivine}^m$). The consequence is a decrease of the solubility constant from acidic to ultrabasic melts.

With regard to point (iii), the solubility is positively correlated with the temperature, because of the expansion of the melt upon heating at constant pressure. Indeed, the higher the temperature the larger the molar volume of the silicate and the lower the entropic cost to insert the noble gas. Though the other terms ($\gamma_m^E$ and $(\rho_m k_B T)^{-1}$) figuring in the expression of the solubility constant S are negatively correlated with the temperature, their contribution is too small to balance the dominant tendency imposed by $\gamma_m^S$. Notice also that the rise of the solubility with T is all the more pronounced as the noble gas is bigger and the melt more depolymerized (see Fig.1). Experimentally the increase of the solubility with T is generally observed at superliquidus temperatures (but not in the glass transition region where a complex behaviour is observed) even if data points are scattered and the interval of temperature is rather limited (Hayatsu and Waboso, 1985; Jambon et al., 1986; Lux, 1987; Carroll and Stolper, 1993; Shibata et al, 1998; Mesko et al., 2002; Marrocchi and Toplis, 2005; Tournour and Shelby, 2008a,b).

## 3.2 Pressure dependence of the solubility

The solubility (in weight fraction) of a noble gas in a silicate melt at given (P,T) conditions was evaluated from the expression (see Eq.2), $X_W = n_g \gamma_m / n_m \gamma_g$ , where $n_g$ and $n_m$ are the densities of the noble gas fluid and of the melt at equilibrium and where $\gamma_m$ and $\gamma_g$ are the solubility parameters of the noble gas in the melt and in its own fluid, respectively. Hence the evolution with pressure of $X_W$ will depend on the pressure dependence of the ratios $n_g/n_m$ and $\gamma_m/\gamma_g$ . For illustration the pressure evolution of all these quantities are presented in Fig.3 when a noble gas fluid (He, Ne, Ar or Xe) is coexisting with a MORB melt at 2273 K. Thus the ratio $n_g/n_m$ increases quasi linearly at low pressures when the fluid is close to ideality and the melt quasi incompressible (then $n_g \sim P/RT$ and $n_m \sim constant$) and next levels off in the kbar range before saturation at higher pressure when the compressibility of the supercritical rare gas fluid becomes similar to that of the silicate melt. On the other hand $\gamma_m$ and $\gamma_g$ both decrease strongly when the pressure increases and yet the ratio $\gamma_m/\gamma_g$ changes only by a factor of two over all the pressure range of investigation (0-200 kbar). This equalization is observed with all noble gases although the heavier the noble gas the stronger the decrease of $\gamma_m$ and $\gamma_g$ with the pressure (for He the latter quantities decrease by two orders of magnitude between 0 and 100 kbar instead of five orders of magnitude for Ar). The consequence of these findings is that the solubility $X_W$ behaves as the ratio $n_g/n_m$ at low pressures and as the ratio $\gamma_m/\gamma_g$ at high pressures (see Fig.3). The same behaviour is observed in rhyolite and in molten olivine (not shown), the only one difference being that the absolute magnitude of $\gamma_m$ at a given pressure depends on the degree of depolymerization of the melt ($\gamma_{olivine} < \gamma_{MORB} < \gamma_{rhyolite}$ as shown in Fig.2 for P~0).

To better understand why $\gamma_m$ and $\gamma_g$ decrease so drastically when the pressure increases it is worthwhile to evaluate their respective entropic ($\gamma^S$) and energetic ($\gamma^E$) contributions ($\gamma = \gamma^S \times \gamma^E$ ). The pressure dependence of $\gamma_m$ , $\gamma_m^S$ and $\gamma_m^E$ for He, Ne, Ar and Xe in MORB melt at 2273K and that of $\gamma_g$ , $\gamma_g^S$ and $\gamma_g^E$ in the corresponding supercritical rare gas fluid is shown in Fig.4. It is clear that the entropic penalty to insert a noble gas either in the melt or in the parent fluid is the dominant contribution to the solubility parameters $\gamma_m$ and $\gamma_g$. This entropic penalty increases strongly with the pressure (i.e. $\gamma_m$ and $\gamma_g$ decrease when P increases) and with the size of the noble gas atom as well. In contrast the pressure dependence of the energetic contributions ($\gamma_g^E$ and $\gamma_m^E$) is much weaker but is more subtle. Thus at low and moderate pressures (P≤20 kbar) it is energetically more favorable to insert in a silicate melt a highly polarizable atom (e.g. Xe) than a weakly polarizable atom (e.g. He) whereas at high pressure (P>30 kbar) the energy gain resulting from the solvation of a highly polarizable atom vanishes because repulsive forces between the solute and the atoms of the melt override the (attractive) dispersion forces. Hence, for He and Ne atoms the energetic contribution $\gamma_m^E$ increases slightly the insertion penalty in the melt at any pressure, whereas for a Xe atom $\gamma_m^E$ reduces somewhat the insertion penalty at low pressures but increase it at high pressures (compare He with Xe in Fig.4). Similar trends are found with $\gamma_g^S$ and $\gamma_g^E$, respectively (see Fig.4).

Coming back to the solubility, our results for He, Ne, Ar and Xe solubility in rhyolite, MORB and molten olivine at 2273 K are presented in Fig.5 as function of the pressure (for a better comparison between noble gases the solubility is expressed in mole fraction in this figure, see Eq.3 for a definition). The solubility of noble gases increases steadily in the low pressure range (0~10 kbar) and then slows down before reaching a maximum value followed by a gradual decrease at high pressures. The solubility maximum tends to shift towards a lower pressure when the size of the noble gas increases, a finding in agreement with the prediction of a statistical theory (Guillot and Sarda, 2006). Notice that the solubility of He and Ne is obtained with a fair accuracy ($\pm$ 10% or less) over the whole pressure range under investigation (0-200 kbar). In contrast, the error bars associated with Ar and Xe solubility grow very significantly above ~100 kbar and ~30 kbar respectively. Concerning the evolution with the melt composition, for a given noble gas the solubility is the highest in the rhyolitic melt, the lowest in molten olivine and is intermediate in the basaltic liquid. This findings are in agreement with noble gas solubility data at low pressure (e.g. in Fig.1) and with Ar solubility data in silicate liquids under pressure (White et al., 1989; Carroll and Stolper, 1993; Bouhifd et al., 2006).

Ar solubility in some melts appears to undergo a relatively sudden decrease when the pressure is raised above a threshold value which depends nontrivially on melt composition. For example, in using a laser heated diamond anvil cell experiments, Chamorro et al. (1996, 1998) and Bouhifd et al. (2006, 2008) have observed an Ar solubility drop in liquid silica and in molten olivine at virtually the same pressure (~50 kbar) although the structure of these melts are completely different from each other (Kohara et al., 2004; Mei et al., 2007) and the Ar contents differ by more than one order of magnitude in these two melts (~5 wt% in silica at 50 kbar as compared with ~0.2 wt% in molten olivine at the same pressure). In Fig.6 are compared the results of our calculations for Ar in molten olivine at 2273K with the Ar solubility data of Chamorro et al. (1998) and Bouhifd and Jephcoat (2006). The agreement is excellent in the pressure range 30-50kbar but the simulation does not reproduce the Ar solubility drop at about 50 kbar. In contrast, the theoretical curve levels off above 50 kbar and shows a slow decrease of the solubility at higher pressure, a behaviour that we also observe in our calculations with He and Ne (but with a better statistics, see Fig.5).

We have evaluated the solubility of Ar in liquid silica at 2600K, a temperature that is comparable with the temperature reached in laser heated diamond anvil cell (Chamorro et al., 1996). In Fig.7 are reported the results of the TPM calculations (see the *full triangles*), the high pressure solubility data of Chamorro et al. (1996), those of Bouhifd et al. (2008) and the low pressure data (0.25-6 kbar) of Walter et al. (2000) obtained with vitreous silica at 1473K. At low pressure, the calculated solubility constant for Ar (~72 $10^{-5}$ $cm^3$ STP $g^{-1}bar^{-1}$) compares well with the solubility constant evaluated by Walter et al. (2000) in vitreous silica ($S_{Ar}^{exp}$~79 $10^{-5}$ $cm^3$ STP $g^{-1}bar^{-1}$). The agreement between simulation and experiment is still excellent up to ~40 kbar, but above this pressure the simulation predicts a steady increase of the Ar solubility up to ~100 kbar and a solubility maximum afterwards, at variance with the sudden solubility drop observed experimentally at about 50 kbar.

A possible origin of this discrepancy could lie in the inaccuracy of the TPM at high Ar contents in the melt (e.g. at ~50 kbar there are ~7 mole% of Ar in liquid silica) because the method evaluates the excess chemical potential of Ar in the melt at infinite dilution. So to test the accuracy of the TPM we have evaluated the Ar solubility by implementing a MD simulation with explicit interface. In this method the noble gas fluid is explicitly in contact with the silicate melt through the presence of a fluid-liquid interface in the simulation box (e.g. Guillot and Sator, 2011). The entire system is equilibrated at fixed T and P, and the noble gas atoms are free to move across the interface separating the supercritical fluid (e.g. Ar) and the liquid silicate. The noble gas solubility is then obtained by counting the average number of noble gas atoms present within the melt in the course of the MD run. In the present case, a supercritical phase composed of 300 Ar atoms is in contact with a silica melt composed of 999 ions (Si and O). Very long simulation runs (up to $10^8$ MD steps or 0.1 $\mu$s) were carried out to reach a sufficient accuracy because of the rather high viscosity of the simulated melt (~20 Pa.s at 2600K). The calculated Ar solubility from this simulation is also shown in Fig.7 (*open triangles*). The error bars associated with each point were determined from a statistical analysis of the MD runs. It is clear that the direct method reproduces quite accurately (within the statistical uncertainties of the two methods) the results obtained by the TPM, and confirms the absence of a solubility drop in the pressure range found experimentally (~50 kbar). Moreover, the compressibility curve of liquid silica (i.e. molar volume versus pressure, not shown) is barely modified by the presence of a significant amount of Ar atoms (e.g. $X_{Ar}$ ~15 mole% at 100 kbar). This suggests that Ar atoms have no specific interactions with the silica network, the latter accommodating the former interstitially (more detailed structural information is provided later). This result explains a posteriori why the infinite dilution approximation used with the TPM works so well in liquid silica even at high Ar contents.

In this context of conflicting results between theory and experiment it is important to know whether the simulation results depend on the potential used to describe liquid silica. To do so, the Ar solubility has been recalculated in using the PPSM potential (Guillot and Sator, 2007a), instead of the CHIK potential (Carré et al., 2008), to simulate the silica melt. However, it is noteworthy that the CHIK potential reproduces the EOS of liquid silica with accuracy because the *exp-6* term (repulsion-dispersion contribution to the pair potential) is truncated at 6.5 A. If the *exp-6* term is not truncated then the density of the simulated melt is too high (e.g. at 2600K and P~0, $n_{sim}$ = 2.34 g/cm$^3$ instead of 2.2 g/cm$^3$ experimentally). This situation is encountered with most of empirical pair potentials for silica (Soules et al., 2011; Guillot and Guissani, 1996a; Vollmayr et al., 1996). Although the truncation of the *exp-6* term has no theoretical background, it is an efficient way to fit the pressure of a simulated silica melt. In the case of the PPSM potential, a cut off distance equal to 6.0 A is required to reproduce with accuracy (i.e. $\pm 1\%$ for the density) the compressibility curve of molten silica (Tsiok et al.,1998). In modeling the silica melt with this truncated PPSM potential, we have re-calculated the Ar solubility at 2600K using the TPM. The calculated Ar solubility curve (not shown in Fig.7 for convenience) is virtually identical to that obtained with the CHIK potential (e.g. at P=82 kbar, $X_{Ar}$ = 7.9$\pm$0.9 wt% with PPSM and 8.1$\pm$0.9 wt% with CHIK). Therefore, the pressure behaviour of the calculated Ar solubility does not depend on the pair potential used to describe molten silica.

Bouhifd et al. (2006, 2008) measured the pressure behaviour of the Ar solubility in various melt compositions (e.g. anorthite, C1-chondrite, and sanidine in addition to olivine and silica) and found that a solubility drop is also observed in those melts. Furthermore, the pressure at which the Ar solubility drop occurs, seems to be correlated with the Al-contents of the melt. In Al-free melts (e.g. olivine and silica) the Ar solubility drop is observed at the same pressure (~50 kbar) whereas in Al-bearing silicate melts this pressure increases with the Al-contents (around 100 kbar in C1-chondrite melt, ~140 kbar in sanidine melt and ~170 kbar in anorthite melt). Thus for a haplogranitic melt the Ar solubility drop is expected at about 120-130 kbar, a pressure range located above the maximum pressure (~80 kbar) investigated by Schmidt and Keppler (2002). For a basaltic melt (e.g. tholeiite) the same correlation leads to a pressure of about 140 kbar when the solubility drop begins.

To check if this latter prediction is supported by calculations, the Ar solubility in MORB at 2273K calculated by the TPM is compared with the available solubility data for Ar in basaltic melts (olivine tholeiite) of White et al. (1989) and Carroll and Stolper (1993) (Fig.8). The latter experimental data only cover the pressure range 2.5-25 kbar. Moreover, the simulated MORB melt is slightly richer in silica than the olivine-tholeiite investigated experimentally (50.6 wt% $SiO_2$ in MORB as compared with 46-49 wt% in olivine-tholeiite) and the simulated temperature is higher than the one at which the experiments were conducted (2273 K instead of 1480-1873K for Carroll and Stolper and 1773-1873K for White et al.). Because of these differences, the Ar solubility is expected to be slightly higher in our simulated melt than in the experimental samples (see Fig.1 for the trends with temperature and melt composition). Therefore, in the pressure range where a comparison is possible (2.5-25 kbar), the agreement between simulation and experimental data is quite satisfactory. At higher pressure, between 30 and 50 kbar, the theoretical curve levels off and shows a downward trend above 80 kbar. Although the error bars associated with the calculated points becomes large above 120 kbar, there is no evidence of a solubility drop up to 150 kbar. To check the robustness of our results obtained with the TPM, we have implemented the direct method with explicit interface, already used with silica, to evaluate the Ar solubility in the MORB melt. Here again very long simulation runs are needed ( ~0.1 $\mu s$ for a system composed of 300 Ar atoms and 1,000 ions) because of the low solubility of Ar in MORB (only ~0.5 mole% at 30 kbar, that corresponds to ~2 Ar atoms on average in a silicate melt composed of 1,000 ions). The results are also reported in Fig.8. The solubility curve is quite close to that calculated with the TPM (compare the dashed curve with the full curve) in spite of a slight downward shift of the Ar solubility obtained by the direct method (this is due to a finite size effect, which is immaterial in the present context). In particular, no solubility drop is observed in the 120-200 kbar pressure range. So, in conclusion, our calculations do not support the prediction of Bouhifd et al. (2006, 2008) of a sudden Ar solubility drop around 140 kbar in basaltic melts.

The fact that the calculated Ar solubility curves in liquid silica, in molten olivine and in a basaltic liquid do not exhibit a solubility drop (or a marked decrease by one order of magnitude) in the pressure range where this feature is observed in some experimental studies (Chamorro et al., 1996, 1998; Bouhifd et al., 2006, 2008) requires comments. First, the experimental situation is not simple because the pressure at which the Ar solubility drop is observed experimentally is virtually the same in liquid silica and in molten olivine (P ~50 kbar), two melts which exhibit very different thermodynamic and structural properties, whereas it is observed at a much higher pressure (~140 kbar) in liquid sanidine, a siliceous melt. Although Bouhifd et al. (2006, 2008) have pointed out a correlation between the threshold pressure of the solubility drop and the Al-contents in the silicate melts they investigated, the role of Al is elusive. These authors argued that with increasing pressure, the fraction of highly coordinated Al in aluminosilicate melts increasing smoothly from [4]Al to [5]Al and [6]Al (Lee et al., 2004, 2006; Allwardt et al., 2005), this could lead to a minimum porosity below which argon atoms can no longer be accommodated by the melt structure. However this scenario is not supported by the present simulation data. Indeed we do observe in our simulated melts a progressive increase of highly coordinated species [5,6]Al (but also [5]Si) with increasing pressure (for details see Guillot and Sator, 2007b), in agreement with structure data, and still the Ar solubility shows no evidence of a marked decrease with pressure, just a gradual decrease (at least for the melt compositions and for the pressure range investigated in the present study).

A source of uncertainty is the force field describing the properties of noble gas fluids in the HT-HP range of investigation. As discussed in the Appendix B, the equation-of-state (EOS) of these fluids at such extreme conditions is not well constrained. For Argon, Ross et al. (1986) have proposed a set of parameters for the (exp-6) model based upon shock wave data and static compression of solid argon. This interaction potential describes a fluid slightly more compressible than the one modeled by the Tang-Toennies potential that we have implemented in our calculations (see Appendix B and Fig.B1). To estimate the influence of the fluid phase properties on the solubility of Ar in silicate melts we have re-evaluated by the TPM the solubility parameter of Ar in its own parent fluid, $\gamma_g^{Ar}$, and introduced the new results in the expression of the solubility (see Eq.2). By comparison with our results obtained with the Tang-Toennies potential, those carried out with the Ross potential lead to an increase of the Ar solubility in silicate melts at 2273 K about 3% at 10 kbar, 15% at 30 kbar, 36% at 50 kbar, 60% at 80 kbar, and ~130% at 120 kbar, whereas the density of fluid Ar in the same conditions increases by ~2.8%, 4.4%, 5.2%, 6.2% and 7.1% , respectively.

## 3.3 Noble gas diffusion in basaltic melts

A precise knowledge of noble gas diffusion in basaltic melts is important to better constrain degassing scenarios of the Earth's mantle. The high $^4$He/$^{40}$Ar ratio measured in the vesicles of MORB glasses is sometimes interpreted as an incomplete degassing of the $CO_2$-bearing melt, where the difference between He and Ar diffusivities is invoked to explain the deficit in Ar atoms with respect to He atoms (Aubaud et al., 2004). More generally it is believed that a kinetic disequilibrium induced by differences in noble gas diffusion may play a role in noble gas fractionation during the ascent and emplacement of magmas.

To evaluate the self diffusion coefficients of He, Ne, Ar, and Xe in a basaltic melt, we have performed MD simulations where one noble gas atom is diluted in a MORB melt composed of 1,000 ions. The system is equilibrated at 20 kbar for comparison with data of the literature. The self diffusion coefficient $D_X$ associated with the noble gas atom $X$ is evaluated from its mean square displacement,

$$D_X = lim_{t \to \infty} \frac{<(r_X(t)-r_X(0))^2>}{6t} \tag{11}$$

where $r_X(t)$ is the position of the noble gas atom in the melt at time, t, and where the bracket expresses an average over many time steps taken as origin times. Simulation runs of 10 ns were carried out to reach a sufficient statistics for $D_X$ (~$\pm$20%). The results are shown in Fig.9 in a $log\ D_X$ versus $1/T$ representation. In the liquid range investigated here and at a given temperature, He has the largest diffusion coefficient followed by Ne, Ar and Xe. This hierarchy is similar to that observed with the solubility. Note that at magmatic temperatures, the differences in the magnitude of $D_X$ between light and heavy noble gases are not very large ( ~ one order of magnitude between He and Ar at 1673K). Moreover the activation energy increases non linearly with the size of the noble gas atom ($D_X = D_0\ e^{-E_a/RT}$ where $E_a$ = 72.4 kJ/mol for He, 105.1 for Ne, 119.7 for Ar and 143.2 for Xe). The calculated diffusion coefficients are close to the experimentally-determined noble gas diffusion data of Lux (1987) for a tholeiitic melt at 1623K (see Fig.9 for a comparison). Nowak et al. (2004) have published Ar diffusion data for synthetic melts of basaltic composition in the temperature range 1623-1773K. These data are also reported in Fig.9 and are smaller than that of Lux (1987) by roughly one order of magnitude. This deviation could originate in part from a difference in composition, the melt investigated by Nowak et al. (a haplobasalt) being slightly more polymerized than the tholeiite of Lux (1987). The calculated diffusion coefficient of molecular $CO_2$ in a MORB melt (Guillot and Sator, 2011) also is similar to that calculated for Ar (Fig.9). This result supports the view of Nowak et al. (2004) that in basaltic melts the diffusion of Ar can be used as a proxy of molecular $CO_2$ (for a discussion of the diffusion of $CO_2$ and $CO_3^{2-}$ species in silicate melts see Nowak et al. (2004) and Guillot and Sator (2011)). Note that the diffusion coefficients of Ar and $CO_2$ are close to each other also in silicic melts (Watson, 1981; Blank et al., 1991; Behrens, 2000, 2010).

## 3.4 Melt structure around noble gases

The structure of silicate melts near dissolved noble gases is still poorly known. We are aware of only one experimental study by X-ray absorption spectroscopy (Wulf et al., 1999) on Kr dissolved in a silica melt at 1473 K and 7 kbar showing a densely packed environment of oxygen atoms around Kr. The implementation of the TPM allows one to evaluate the pair distribution functions, $g_{X-S}$ (r), between the noble gas atom $X$ and the ionic species $S$ (where $S$ = O, Si, Al, Ca,...) in the silicate melt (see Fig.A3 in Appendix A). The distance $R_{X-S}$, which corresponds to the first maximum of the pair distribution function $g_{X-S}(r)$, is the most probable distance between the noble gas atom and the nearest neighbor ion of species $S$. This distance is reported in Fig.10 for He, Ne, Ar and Xe in the three simulated melts (at T=2273K and P~0) and is represented as function of the Lennard-Jones diameter $\sigma_{X-S}$ associated with the corresponding pair potential $u^{LJ}_{X-S}(r)$ (see Eq.B3 in Appendix B). The line drawn in Fig.10 expresses the equality $R_{X-S} = \sigma_{X-S}$. If for a pair *(X,S)* the distance $R_{X-S}$ is equal or smaller than $\sigma_{X-S}$ (the corresponding data point is then located below the line drawn in Fig.10), the interaction between the noble gas atom and the first neighbor ion of species $S$ is repulsive on average and the two atoms are in close contact with each other. On the other hand, if $R_{X-S} > \sigma_{X-S}$ (the data point is located above the line), the noble gas atom is not in close contact with the ion. A rapid analysis of Fig.10 shows that in the three melts, noble gas atoms are not in contact with network former ions (Si, Al and Ti) neither with $Fe^{3+}$, considered sometimes as a network former cation (Mysen, 2006). This finding is expected because these cations are embedded (on average) into their oxygen coordination shell ($TO_4$) and noble gas atoms cannot penetrate into these densely packed regions. Accordingly, the noble gases are found to be in close contact with the oxygens ($R_{X-O} \leq \sigma_{X-O}$ in Fig.10).

Wulf et al. (1999) derived from X-ray absorption spectra an average distance about 3.45 A between Kr and oxygen atoms in a silica melt at 1473 K , a value which is identical to the one we obtain for the distance $R_{Kr-O}$ in liquid rhyolite at 2273 K (Kr is not shown in Fig.10). Furthermore, Zhang et al. (2009) have evaluated by *ab initio* MD calculation the distance $R_{X-O}$ for the noble gases diluted in liquid silica. Their results are quite close to our own evaluations in liquid rhyolite (e.g. $R_{He-O}$ = 2.63 A in silica and 2.65 A in rhyolite; $R_{Ne-O}$ = 2.81 A in silica and 2.75 A in rhyolite; $R_{Ar-O}$ = 3.26 A in silica and 3.25 A in rhyolite; $R_{Kr-O}$ = 3.31 A in silica and 3.45 A in rhyolite; $R_{Xe-O}$ = 3.53 A in silica and 3.55 A in rhyolite).

In contrast, the most probable distances between noble gases and network modifier cations ($Fe^{2+}$, Mg, Ca, Na and K) exhibit a more complex pattern (Fig.10). When a noble gas atom is in the neighborhood of an alkali ion (Na or K) the most probable location is to be in close contact with it (e.g. $R_{X-Na} \leq \sigma_{X-Na}$). In the case of Ca, the distance $R_{X-Ca}$ is slightly greater than the corresponding L-J diameter (by $\sim$ +0.1-0.2 A) for all noble gases except He, for which $R_{He-Ca} \sim \sigma_{He-Ca}$ . The case of $Fe^{2+}$ looks like that of Ca but with a more pronounced evolution of $R_{X-Fe^{2+}}$ with the size of the noble gas (the larger the noble gas the greater the deviation between $R_{X-Fe^{2+}}$ and $\sigma_{X-Fe^{2+}}$). With regard to Mg, the most probable location of the noble gas depends more clearly on its size. Whereas He and Ne are characterized by a value of $R_{X-Mg}$ greater than $\sigma_{X-Mg}$ by about +0.3-0.4 A, Ar and Xe are located at a much greater distance from Mg ($\sigma_{X-Mg} \sim$ +0.6-0.8 A). So, Ar and Xe are never located in the immediate vicinity of a Mg cation because the oxygen coordination shell is tightly bound to the latter one (Mg is n-coordinated with n=4, 5 and 6, the respective weight of these contributions depending on melt composition, see Shimoda et al. (2007), Neuville et al. (2008), Wilding et al (2008), and Guignard and Cormier (2008)). It is also apparent that the distances $R_{X-S}$ generally shrink a little bit in going from the most polymerized melt (rhyolite) to the most depolymerized one (molten olivine). This trend is correlated with the free volume accessible into the melt, the more depolymerized the melt the smaller the free volume (and the lower the noble gas solubility).

We will focus our structure analysis on noble gas-oxygen correlations because oxygen is the dominant component in silicate melts with an atomic fraction which does not depend very much on melt composition (~63% in rhyolite, ~61% in MORB and ~57% in olivine) and because the noble gas atoms interact mainly with oxygen. In Fig.11 the noble gas-oxygen pair distribution functions (PDF) $g_{X-O}(r)$ are shown in the three melts at 2273 K and P = 0 and 30 kbar. A striking difference appears at P=0 kbar between the PDF of noble gases in rhyolite and those in MORB and olivine. In rhyolite the first maximum of $g_{X-O}(r)$ is <1 for Ar and Xe and barely reaches one for He and Ne. This feature means that the first oxygen shell around the noble gas is depleted in oxygen atoms with respect to a continuous medium (and the larger the noble gas atom the stronger the depletion). In contrast, in MORB and olivine melt the intensity of the first maximum of $g_{X-O}(r)$ is >1, meaning that the first solvation shell is well developed. However, over the extent of the first solvation shell the difference between a depleted shell and a well developed shell is only a few oxygen atoms. This is illustrated in Fig.12 where the number of oxygen atoms, $N_O$, in the first solvation shell is shown as function of melt composition and pressure. At P = 0 kbar there are ~8.5 oxygen atoms around Xe in rhyolite, ~12 in MORB and almost 14 in molten olivine, whereas around He these numbers are as low as ~5.5, 7.5 and 8 respectively. These numbers suggest that the free volume accessible to a noble gas atom is larger in rhyolite than in depolymerized melts as MORB and olivine. Indeed, the calculated accessible free volume for He is important and decreases moderately with the degree of depolymerization of the melt (~30% in rhyolite, ~25% in MORB and ~21% in olivine) whereas for Xe the free volume is very small and varies strongly with melt composition (~1.5% in rhyolite, ~0.3% in MORB and ~0.1% in olivine).

Our values for noble gas coordination numbers in rhyolite under pressure are in accordance with those evaluated in liquid silica by Zhang et al. (2009) in using *ab initio* MD simulation. The evolution of $N_O$ with pressure is similar in both melts with a rapid rise between 0 and 30 kbar, followed by a leveling off at higher pressure (compare Fig.12 with Fig.4 of Zhang et al. (2009) in noticing that the molar volume of liquid silica can be converted in pressure by using the EOS of Karki et al. (2007)).

When increasing pressure, the magnitude of the first peak of $g_{X-O}(r)$ increases and its position shifts to slightly lower r values (see Fig.11). These modifications are pronounced between 0 and 30 kbar (especially in rhyolite) and become barely visible at a higher pressure (not shown). Correspondingly, the number of oxygen atoms, $N_O$, surrounding a noble gas atom increases rapidly in the 0-30 kbar pressure range (the increase is more marked in rhyolite than in the other two melts) goes through a weak maximum (with Ar and Xe) and reaches a plateau value at higher pressure. The coordination change of Si and Al atoms in this pressure range is too small (less than 10% for Al and a few percent for Si; see Lee et al. (2004), Allwardt et al. (2005), and Gaudio et al. (2008)) to play a significant role in the initial increase of $N_O$ with pressure. However, the latter one is correlated to the decrease of the T-O-T intertetrahedral angle (where T= Si and Al). This angle compression (El'kin et al., 2002; Clark et al., 2004; Malfait et al., 2008) is the more effective the higher the degree of polymerization of the melt (Guillot and Sator, 2007b; Fig.11). As for the absolute magnitude of $N_O$ at high pressure, it depends on the size of the noble gas (the larger the size the higher $N_O$) and marginally on the melt composition. In contrast, the free volume accessible to noble gas atoms decreases steadily with the pressure and is lower when the melt is less polymerized (see Fig.13). In fact it is the population of cavities capable of accommodating the noble gas that diminishes with the pressure, whereas the number of oxygen atoms that make up these cavities is roughly the same regardless of the composition of the melt.

**Appendix A**

The statistical uncertainties associated with the calculation of the noble gas solubility by the TPM (Widom, 1963) are estimated in the following way. The solubility parameter $\gamma$ of a noble gas in a solvent $i$ (silicate melt or parent fluid) is given by,

$$\gamma_i = e^{-\mu_i^{ex}/k_B T} = <e^{-\psi/k_B T}>_N = \frac{1}{N_{MD}} \sum_{l=1}^{N_{MD}} (\overline{e^{-\psi/k_B T}})_l \tag{A1}$$

where $\psi$ is the interaction energy between the test particle (the noble gas atom) and the solvent, $N_{MD}$ is the total number of MD steps sampled by the TPM and $(\overline{e^{-\psi/k_B T}})_l$ an average evaluated on the MD step $l$. The latter averaging is obtained from

$$\overline{e^{-\psi/k_B T}} = \frac{1}{N} \sum_{j=1}^{N_u} e^{-\psi_j/k_B T} \tag{A2}$$

where $N$ is the number of cubelets used to describe the volume of the simulation box (i.e. $41^3$ or $161^3$), $N_u$ the number of unoccupied cubelets and $\psi_j$ is the interaction energy between the test particle inserted at the position $j$ and the solvent particles. To reach an accurate value for $\overline{e^{-\psi/k_B T}}$ one only needs to check the convergence of the result when increasing the number of cubelets. With $N = 41^3$ or $161^3$ one can locate all cavities present in the solvent configuration that can accommodate a noble gas atom. In contrast, because of the density fluctuations exhibited by the solvent, the step value $(\overline{e^{-\psi/k_B T}})_l$ may vary significantly from one MD step to another, this feature being all the more pronounced when the investigated pressure is high (the denser the solvent) and the noble gas is large.

The evolution as function of running time of the solubility parameter $\gamma_m$ for He, Ne, Ar and Xe in a MORB at 50 kbar and 2273 K is illustrated in Fig.A1. When the fluctuations of $\gamma_m$ are barely perceptible for He and Ne, they remain visible for Ar and very significant for Xe. So the occurence of cavities large enough to accommodate an atom as big as Xe in a MORB melt at 50 kbar is a relatively rare event that requires a very long MD run (e.g. 10 ns) to be sampled with a reasonable accuracy. To estimate the statistical uncertainties on $\gamma$ we have evaluated the variance $<(\overline{\gamma} - \gamma)^2>$ where $\overline{\gamma}$ is the average given by Eq.A2 and where the averaging $<\cdots>$ is taken over the number of sampled MD steps (see Eq.A1). In Fig.A2 is reported the evolution of the ratio $\Delta\gamma_m/\gamma_m$ (where $\Delta\gamma_m$ is the square root of the variance) for He, Ne, Ar and Xe in a MORB melt at 2273K as function of the pressure. The statistical uncertainties increase with the pressure and more so with increasing size of the noble gas. Thus the uncertainties for Xe increase from a few percent at low pressures to ~100% at 100 kbar, they still amount to ~ 40% for Ar at this very pressure but are below 10% for He and Ne at 200 kbar. These uncertainties are weakly dependent on melt composition, the higher the degree of polymerization of the melt the smaller the error bars (not shown). As for the uncertainties on the evaluation of $\gamma_g$, the solubility parameter of a noble gas in its own fluid, they are generally smaller by roughly one order of magnitude compared with those associated with $\gamma_m$ at the same thermodynamic conditions (e.g. in Fig.A2 for Ar).

The statistical uncertainties associated with the evaluation of the solubility, $X_W$, arise from the cumulative errors in estimating successively $n_g$, $n_m$, $\gamma_g$ and $\gamma_m$. As the densities, $n_g$ and $n_m$, of the fluid phase and of the melt are evaluated with $\pm$ 1% accuracy, the numerical uncertainties on $X_W$ are dominated by the statistics of $\gamma_m$ and $\gamma_g$ and specially by that of $\gamma_m$ for Ar and Xe at high pressure. In the following the error bars are estimated by adding all these uncertainties.

Another way to estimate the accuracy of the TPM is to evaluate the pair distribution functions (PDF) between a noble gas atom and an element of the silicate melt (e.g. *Si, O, ...*) or an atom of the coexisting noble gas fluid. These functions are calculated in two ways, on the one hand by MD simulation where a noble gas atom is diluted in the silicate melt and on the other hand by the TPM. The following definitions are used for calculation,

$$g_{X-S}^{MD}(\mathrm{r}) = \frac{1}{N_s} < \sum_{i=1}^{N_s} \delta(r - r_{X-i}) >_{N+1} \tag{A3}$$

$$g_{X-S}^{TPM}(\mathrm{r}) = \frac{1}{N_s} < \sum_{i=1}^{N_s} \delta(r - r_{X-i}) \ e^{-\psi/k_B T} >_N / < e^{-\psi/k_B T} >_N \tag{A4}$$

where X is the solute atom (e.g. $X = Ar$), $N_S$ is the number of atoms of species S in the melt (e.g. $S = O$) or in the fluid phase (e.g. $S = Ar$), $r_{X-i}$ is the distance between the atom X and the i$^{th}$ atom of species $S$, $< \cdots >_N$ indicates an average performed over the configurations of the N particles of the simulated pure solvent and $< \cdots >_{N+1}$ an average over the *(N+1)* atoms of the simulated mixture. For illustration the results of these two evaluations are compared in Fig.A3 for $g_{X-O}$(r) and $g_{X-Si}$(r) (with X= He, Ar and Xe) in a MORB melt at 2273K and 20 kbar. The good agreement between the two calculations (TPM versus MD) at 20 kbar implies that the TPM samples accurately the structure of the silicate melt at these conditions. At higher pressures, and mainly for Ar and Xe, the PDFs evaluated by the TPM become noisy (see Xe in MORB at 100 kbar in Fig.A3), which indicates a less accurate sampling. A systematic investigation of the PDFs shows that the TPM leads to an accurate sampling of the melt structure as long as the pressure does not exceed ~50 kbar for Xe in MORB, and ~150 kbar for Ar in MORB (for He and Ne the sampling is accurate up to 200 kbar at least). This conclusion is in accordance with the increased variance of $\gamma_m$ discussed above. A better sampling at high pressures would require so much increase of the MD runs that it is untractable with present day computational resources.

**Appendix B**

*Noble gas-noble gas interaction potentials*

To simulate rare gas fluids we have used the van der Waals potentials of Tang and Toennies (2003), which reproduce very accurately *ab initio* calculations and the best fitted empirical potentials for homogeneous rare gas dimers. For instance, the Tang-Toennies (T-T) potentials are used as benchmarks to test van der Waals interactions in density-functional theory (Kannemann and Becke, 2009). However, the analytical expression of the T-T model for a rare gas dimer is not that simple because the potential energy curve requires five parameters to be described. For reason of compatibility with our MD code, we have used a simpler analytical form for the potential energy v($r$), namely the well known (exp-6) model,

$$v(r) = A\, e^{-r/\rho} - C/r^6 \qquad\qquad (B1)$$

and adjusted the three parameters $A$, $\rho$ and $C$ to reproduce the original potential energy curves. This analytical form leads to potential energy curves which are virtually undistinguishable from the original T-T model for interatomic separation smaller or equal to the van der Waals distance (a distance corresponding to the potential well minimum), whereas at larger distances a very small deviation occurs, but this is immaterial in the present context (at the high temperature-high pressure investigated here, this is the short range part of the potential that really matters).

To check the ability of these potentials to reproduce the (P,V,T) properties of rare gas fluids, we have evaluated by MD simulation the equation of state (EOS) of He, Ne, Ar and Xe along the isotherm T=300 K. The results are presented in Fig.B1 (see the insert) and are compared with the refractive index data of Dewaele et al. (2003) for He (0.8≤P≤115 kbar) and Ne (7≤P≤47 kbar) and with the EOS of Tegeler et al. (1999) for Ar up to 10 kbar. The agreement is very good for the three noble gases even if the (exp-6) model seems to underestimate slightly the molar volume of He but the way this latter value is deduced from the refractive index is not free from uncertainties (see Dewaele et al., 2003). As for Xe, the 300 K isotherm covers a small range of pressure (up to ~4.5 kbar before cristallization) and the agreement of our MD data with the EOS of Sifner and Klomfar (1994) in this pressure range is also quite satisfying (not shown). Unfortunately, for the (T,P) range investigated in the present study (1673-2600 K and 0-200 kbar) there are no data available in the literature which can be compared directly with our simulation results (see Fig.B1 for the isotherm T=2273 K). However, shock wave data are available for He (Nellis et al., 1984), Ar (van Thiel and Alder, 1966; Ross et al., 1979; Grigoryev et al., 1985; Arinin et al., 2008) and Xe (Keeler et al., 1965; Nellis et al., 1982; Root et al., 2010), and (exp-6) potentials have been fitted to reproduce some Hugoniot data for Ar (Ross et al., 1986) and for Xe (Ross and McMahan, 1980). Moreover, the melting curves of He, Ne, Ar, and Xe (Boehler et al., 2001) have been investigated by statistical models and by computer simulations in implementing the aforementioned (exp-6) interatomic potentials (Young et al., 1981; Ross et al., 1986; Belonoshko et al., 2001, 2002; Saija and Prestipino, 2005, Koči et al., 2007). However values of the parameters A, $\rho$, and C associated with these potentials differ somewhat from those that we have deduced from the T-T model. In general softer core repulsion at high energy is the result because they were fitted on shock wave data which probe highly repulsive configurations occurring at very high pressures (up to the megabar range). At lower pressures, these potentials become less accurate because they do not fit closely the extended region around the potential well minimum, which is accurately known from *ab initio* calculations. So, to evaluate the influence of the rare gas pair potential on our solubility calculations, we have also performed MD simulations with the potential parameters determined by Ross et al. (1986) from shock wave data for Ar. The isotherm 2273 K deduced from this potential is compared in Fig.B1 with the isotherm obtained with the T-T model. The isotherm associated with the Ross potential describes a fluid slightly more compressible than the one simulated with the T-T model (e.g. $\Delta V/V \sim$ - 3% at 30 kbar and $\sim$ -8% at 150 kbar).

*Pair potential for silicate melts*

To simulate fused silica we have implemented the CHIK potential (Carré et al., 2008), which is a pairwise additive potential composed of an electrostatic contribution plus an (exp-6) term for repulsion-dispersion interactions,

$$v_{ij}(r) = \frac{z_i z_j}{r} + A_{ij}\, e^{-r/\rho_{ij}} - C_{ij}/r^6 \qquad\qquad (B2)$$

where $r$ is the distance between the two atoms $i$ and $j$ (Si or O), $z_i$ is the effective charge of the atom $i$ ($z_{Si}$ = +1.9104 e) and $A_{ij}$, $\rho_{ij}$ and $C_{ij}$ are parameters associated with the pair ($ij$). The potential parameters were obtained from a fitting scheme based upon *ab initio* MD calculations. In the present context, the main advantage of this potential is that it reproduces the EOS of silica well (see Horbach, 2008), although this agreement is obtained in an ad hoc way by the truncation of the (exp-6) term at 6.5 A (see text). Moreover this potential was also used by Zhang et al. (2010) in their simulation study of Ar solubility in molten silica.

For rhyolite, MORB, San Carlos olivine, and enstatite we have used the effective pair potential for silicate melts (PPSM) from Guillot and Sator (2007a). The EOS and the structure of liquid silicates of various composition (felsic to ultramafic) are well reproduced by this model potential over a large pressure range (0 ~ 250 kbar). For instance the density of our simulated rhyolitic liquid at atmospheric pressure (2.33 g/cm3 at 1673 K and 2.22 g/cm3 at 2273 K) agrees within 2% with values deduced from the multicomponent mixing models of the literature (Lange and Carmichaël, 1987; Ghiorso and Kress, 2004; Fluegel et al., 2008). Although an EOS for rhyolitic liquids covering the high-pressure range is not yet available, silicic liquids are characterized by a low value of the bulk modulus at atmospheric pressure as compared with basic and ultrabasic liquids ($K_0$~130-160 kbar instead of 200-250 kbar see Rivers and Carmichaël, 1987; Lange, 2003; Liu et al., 2006; Lange, 2007; Tenner et al., 2007; Ai and Lange, 2008; Kuryaeva and Surkov, 2010), a feature which is well reproduced by simulation data (see Table 4 in Guillot and Sator, 2007b). For MORB, the density of the simulated melt is in excellent agreement with the data of Ohtani and Maeda (2001) obtained by sink/float experiment (e.g. along the isotherm 2273 K; at P ~0, $n^{sim}$ = 2.52 g/cm$^3$, $n^{exp}$ = 2.55 g/cm$^3$; at 100 kbar, $n^{sim}$ = 3.32 g/cm$^3$, $n^{exp}$ = 3.30 g/cm$^3$; at 150 kbar, $n^{sim}$ = 3.54 g/cm$^3$, $n^{exp}$ = 3.57 g/cm$^3$). As for molten olivine, although its EOS is not available, we anticipate that the simulated EOS is realistic because with this EOS it is possible to reproduce quite satisfactorily by MD simulation the isothermal compression curves of two other magnesium-rich liquid silicates of peridotitic and komatiitic composition, respectively (for a comparison with experimental data on these liquids see Figs.4-5 in Guillot and Sator, 2007b). Note also that the recent density data on liquid peridotite in the P-T range 7-22 kbar and 2100-2300 K obtained by Sakamaki et al. (2010) by X-ray absorption method are reproduced by the simulation with an accuracy better than 2%.

*Noble gas-silicate interaction potentials*

Due to a lack of experimental and theoretical data the noble gas-silicate interactions are poorly known (however, see Kiselev et al., 1985; Pellenq and Nicholson, 1994; Macedonia et al., 2000; Du et al., 2008). We developed, therefore, new potentials based upon the L-J model which writes,

$$u^{LJ}_{ij}(r) = 4\epsilon_{ij} \left[ \left( \frac{\sigma_{ij}}{r} \right)^{12} - \left( \frac{\sigma_{ij}}{r} \right)^{6} \right] \tag{B3}$$

where $i$ is a noble gas of species $i$ ($i$ = He, Ne,..), $j$ is an ion of species $j$ ( $j$ = O, Si, Al,...), $r$ their separation and $\epsilon_{ij}$ and $\sigma_{ij}$ are parameters associated with the pair ($i,j$). These parameters are determined for each noble gas-ion pair in the following way.

The attractive part of the L-J potential corresponds to the dispersion interaction, $-C_{ij}/r^6$ with $C_{ij} = 4\epsilon_{ij}\sigma_{ij}^6$, and where $C_{ij}$ is the dipole-dipole London dispersion coefficient. It can be evaluated accurately with the Slater-Kirkwood formula (Koutselos and Mason, 1986),

$$C_{ij} = \frac{3}{2} \frac{\alpha_i \alpha_j}{(\alpha_i/N_i^{eff})^{1/2} + (\alpha_j/N_j^{eff})^{1/2}} e^2 a_0^5 \tag{B4}$$

where $\alpha_i$ and $\alpha_j$ are dipole polarizabilities of the atom $i$ interacting with the ion $j$, where $N_i^{eff}$ and $N_j^{eff}$ are the effective numbers of electrons associated with the atom $i$ and with the ion $j$, and where $e$ and $a_0$ are the electron charge and the Bohr radius, respectively. For noble gases their dipole polarizability is well known and their effective number of electrons has been evaluated from *ab initio* calculations (Pyper, 1986). For the ionic species in the silicate melt (O, Si, Ti, Al, Fe³⁺, Fe²⁺, Mg, Ca, Na and K) we have used the dielectric polarizabilities by Shannon (1993). These polarizabilities were obtained by applying the Clausius-Mosotti relation (Lasaga and Cygan, 1982) to a large set of dieletric constant data for various oxides and silicate minerals with the assumption that the polarizability of a complex substance (e.g. a silicate) can be expressed as a sum of the constituent ion polarizabilities (for a review see Shannon and Fisher, 2006). It is only recently that it has become possible to evaluate ionic polarizabilities in condensed phase by *ab initio* calculation (Heaton et al., 2006). For instance the distribution of the oxide anion polarizability in silicate melts of various composition (silica, rhyolite, basalt and enstatite) is rather broad with a most probable value about 1.5-2.0 $A^3$ (Salanne et al., 2008), a value in accordance with the empirical determination of Shannon (~2.0 $A^3$). Because *ab initio* calculated cationic polarizabilities in silicates are not yet available we proceeded with Eq.(B4) by introducing the polarizability values determined by Shannon (1993). Moreover, for the effective numbers of electrons associated with ionic species we used the values of Grimes and Grimes (1997, 1998) with the help of a quantum mechanically-based equation that relates ionic polarizability and effective number of electrons (see Table 1 in Grimes and Grimes, 1998). It is then possible to determine for any noble gas-ion pair the dispersion coefficient $C_{ij}$ given by Eq.(B4), the product of the L-J parameters, $4\epsilon_{ij}\sigma_{ij}^6$, is then fixed for the corresponding pair. To evaluate unambiguously the two parameters $\epsilon_{ij}$ and $\sigma_{ij}$ we need an additional constraint. To do so it is assumed that the sum $\ell_{ij}$ of the van der Waals radius ($r_i^{vdW}$) of the noble gas $i$ and the ionic radius ($r_j^0$) of the ion $j$ in the melt (i.e. $\ell_{ij} = r_i^{vdW} + r_j^0$) is given by,

$$\ell_{ij} = \sigma_{ij} \left[(1 + x (2^{1/6} - 1)\right] \tag{B5}$$

where $0 \leq x \leq 1$ is an adjustable parameter. With the above relationship the value of $\ell_{ij}$ is between $\sigma_{ij}$, (when $x=0$) the distance for which the L-J potential goes through zero, and $2^{1/6} \sigma_{ij}$ (when $x=1$) where the L-J potential reaches its minimum value - $\epsilon_{ij}$. The van der Waals radii of the noble gases are those defined by the Tang-Toennies potentials (2003) that we implemented in our simulations (i.e. 1.49 A for He; 1.54 A for Ne; 1.88 A for Ar; 2.00 A for Kr; 2.18 A for Xe and 2.24 A for Rn), whereas the ionic radii were taken from the reference study of Shannon (1976) namely: $r_{O^{2-}} = 1.36$ A, $r_{Si^{4+}} = 0.26$ A, $r_{Ti^{4+}} = 0.56$ A, $r_{Al^{3+}} = 0.39$ A, $r_{Fe^{3+}} = 0.49$ A, $r_{Fe^{2+}} = 0.70$ A, $r_{Mg^{2+}} = 0.65$ A, $r_{Ca^{2+}} = 1.05$ A, $r_{Na^+} = 1.14$ A and $r_{K^+} = 1.61$ A. It is remarkable that these effective ionic radii lead to cation-oxygen distances ($r_{X-O}$) in silicate melts which coincide almost exactly (within 1%) with the first maximum of the cation-oxygen pair distribution functions obtained by MD simulation (compare values of $r_{X-O}$ given in Table 4 of Guillot and Sator (2007a) with $r_{X-O} = r_X + r_{O^{2-}}$ given by Shannon (1976)).

The final step is to fix the adjustable parameter $x$, after which the L-J parameter $\sigma_{ij}$ of the pair (i,j) is obtained straightforwardly from Eq.B5 and the parameter $\epsilon_{ij}$ deduced from the expression $\epsilon_{ij} = C_{ij}/4\sigma_{ij}^6$ where $C_{ij}$ is evaluated from Eq.B4. To determine the adjustable parameter $x$ we have evaluated the solubility constant S (see Eq.4) of noble gases in a MORB melt at 1673K in using the TPM as it is described in Appendix A. We have first assigned to $x$ the two extreme values, $x=0$ and $x=1$, and compared the results with the noble gas solubility data in tholeiitic basalt melts obtained by Jambon et al. (1986) and Lux (1987). With $x=0$ the calculated solubilities are too large by an order of magnitude for Xe and a factor of two for He and Ne, whereas with $x=1$ the calculated solubilities are too small by an order of magnitude for Xe and much less than an order of magnitude for lighter noble gases. It is rewarding to bracket so closely the solubility data for all noble gases in varying the value of the L-J parameter $\sigma_{ij}$ by only ~10%. By tuning the value of $x$ to 0.25 there is excellent agreement with the solubility data (e.g. in MORB at 1673K the solubility constants expressed in $10^{-5}$ cm$^3$ STP g$^{-1}$bar$^{-1}$ are, $S_{He}$= 59., $S_{Ne}$=38 , $S_{Ar}$=5.1 , $S_{Kr}$=2.7 , $S_{Xe}$= 0.9 as compared with 56.5, 25, 5.9, 3.0 and 1.7 measured by Jambon et al. (1986) and 64, 35, 8.7, 6.3 and 2.7 measured by Lux (1987), for a more detailed comparison with solubility data see Fig.1). The final set of L-J parameters corresponding to $x = 0.25$ is given in Table 2. With the example of silica, it is shown in section 3.2 that the noble gas-silicate potential is transferable and can be used with any silicate-silicate potential as long as the latter one reproduces with accuracy the EOS and the structure of the silicate melt under consideration.

**Appendix C**

**Supplementary data**

Solubility (in $10^{-5}$ cm$^3$ STP g$^{-1}$bar$^{-1}$) of He, Ne, Ar, Kr, Xe, and Rn in simulated silicate melts at P~0.

| silicate | T(K) | $S_{He}$ | $S_{Ne}$ | $S_{Ar}$ | $S_{Kr}$ | $S_{Xe}$ | $S_{Rn}$ |
|---|---|---|---|---|---|---|---|
| silica | 2600 | | | 72. | | | |
| rhyolite | 1673 | 186. | 136. | 35.3 | | 10.2 | |
| | 2273 | 269. | 199. | 65.8 | 46.2 | 25.3 | 21.1 |
| MORB | 1673 | 59.6 | 38.2 | 5.1 | 2.7 | 0.89 | 0.64 |
| | 2273 | 114.1 | 76.6 | 15.3 | 8.9 | 3.5 | 2.6 |
| olivine | 1673 | 18.8 | 10.1 | 0.61 | 0.22 | 0.04 | 0.02 |
| | 1873 | 30.5 | 17.3 | 1.5 | 0.63 | 0.15 | 0.10 |
| | 2273 | 64.8 | 39.7 | 5.2 | 2.5 | 0.71 | 0.47 |
| enstatite | 1673 | 19.9 | 10.0 | 0.55 | | 0.02 | |
| | 1873 | 31.5 | 16.9 | 1.2 | | 0.07 | |

**Table 1**

Potential parameters for noble gases (see Appendix B for details).

| Noble gas | A (kJ/mol) | $\rho$ (A) | C (A$^6$ kJ/mol) |
|-----------|------------|------------|-------------------|
| He | 132917.0 | 0.2051 | 109.84 |
| Ne | 684325.4 | 0.2083 | 523.19 |
| Ar | 2947863.0 | 0.2485 | 5607.64 |
| Kr | 3658508.8 | 0.2673 | 11619.95 |
| Xe | 4474435.5 | 0.2940 | 27142.12 |
| Rn | 25004198.0 | 0.2742 | 42953.34 |

**Table 2**

Lennard-Jones potential parameters for noble gas - silicate interactions (see Appendix B for details).

| | $\varepsilon_{He}$(kJ/mol) | $\sigma_{He}$(A) | $\varepsilon_{Ne}$(kJ/mol) | $\sigma_{Ne}$(A) | $\varepsilon_{Ar}$(kJ/mol) | $\sigma_{Ar}$(A) | $\varepsilon_{Kr}$(kJ/mol) | $\sigma_{Kr}$(A) | $\varepsilon_{Xe}$(kJ/mol) | $\sigma_{Xe}$(A) | $\varepsilon_{Rn}$(kJ/mol) | $\sigma_{Rn}$(A) |
|---|---|---|---|---|---|---|---|---|---|---|---|---|
| O | 0.364 | 2.765 | 0.691 | 2.814 | 1.179 | 3.144 | 1.352 | 3.260 | 1.447 | 3.435 | 1.514 | 3.493 |
| Si | 4.217 | 1.698 | 7.820 | 1.747 | 8.355 | 2.076 | 8.427 | 2.193 | 7.531 | 2.368 | 7.458 | 2.426 |
| Ti | 4.785 | 1.989 | 8.937 | 2.038 | 11.407 | 2.368 | 12.062 | 2.484 | 11.512 | 2.659 | 11.628 | 2.717 |
| Al | 2.269 | 1.824 | 4.206 | 1.873 | 4.927 | 2.203 | 5.089 | 2.319 | 4.703 | 2.494 | 4.704 | 2.552 |
| $Fe^{3+}$ | 4.493 | 1.921 | 8.325 | 1.970 | 10.394 | 2.300 | 10.915 | 2.416 | 10.327 | 2.591 | 10.403 | 2.649 |
| $Fe^{2+}$ | 2.207 | 2.125 | 4.113 | 2.173 | 5.683 | 2.503 | 6.137 | 2.620 | 6.036 | 2.794 | 6.153 | 2.853 |
| Mg | 1.554 | 2.076 | 2.896 | 2.125 | 3.897 | 2.455 | 4.178 | 2.571 | 4.068 | 2.746 | 4.134 | 2.804 |
| Ca | 1.142 | 2.465 | 2.142 | 2.513 | 3.389 | 2.843 | 3.800 | 2.959 | 3.943 | 3.134 | 4.086 | 3.192 |
| Na | 0.523 | 2.552 | 0.984 | 2.600 | 1.596 | 2.930 | 1.804 | 3.047 | 1.891 | 3.221 | 1.966 | 3.280 |
| K | 0.351 | 3.008 | 0.663 | 3.056 | 1.232 | 3.386 | 1.446 | 3.503 | 1.603 | 3.677 | 1.696 | 3.736 |

**Figure captions**

**Fig.1** Solubility of He, Ne, Ar and Xe in rhyolite, MORB, San Carlos olivine and enstatite melts as function of temperature. At 1673K and 1873K molten olivine is in the supercooled state but this does not matter for computational purposes. Calculated values are represented by *full symbols* with guidelines (*circle* = rhyolite, *square* = MORB, *triangle* = olivine, *cross* = enstatite) and the experimental data by *empty symbols* (*circle* = rhyolite, *square* = MORB, *triangle* = diopside and enstatite). References of the experimental data are the following: He/rhyolite (Mesko and Shelby, 2002; Tournour and Shelby, 2008a), He/MORB (Jambon et al., 1986; Lux, 1987), He/enstatite (Kirsten, 1969), He/diopside (Marrochi and Toplis, 2005), Ne/rhyolite (Roselieb et al., 1992; Shibata et al., 1998; Iacono-Marziano et al., 2010), Ne/MORB (Hayatsu and Waboso, 1985; Jambon et al., 1986; Lux, 1987; Miyazaki et al., 2004; Iacono-Marziano et al., 2010), Ne/enstatite (Kirsten, 1969; Shibata et al., 1998), Ar/rhyolite (Roselieb et al., 1992; Carroll and Stolper, 1993; Shibata et al., 1998; Marrochi and Toplis, 2005; Iacono-Marziano et al., 2010), Ar/MORB (Hayatsu and Waboso, 1985; Jambon et al., 1986; Lux, 1987; Carroll and Stolper, 1993; Miyazaki et al., 2004; Iacono-Marziano et al., 2010), Ar/enstatite (Kirsten, 1969; Shibata et al., 1998), Ar/diopside (Marrochi and Toplis, 2005), Xe/rhyolite (Shibata et al., 1998), Xe/MORB (Jambon et al., 1986; Lux, 1987), Xe/enstatite (Shibata et al., 1998). Note that the calculated solubilities of He, Ne and Ar are practically identical in olivine and enstatite melts whereas the solubility of Xe is smaller in enstatite than in olivine melt.

**Fig.2** Solubility parameter $\gamma = \gamma^S \times \gamma^E$ as function of the size of the noble gas in rhyolite, MORB, and San Carlos olivine melts at 2273 K and P~0 (for details see text). The entropic contribution ($\gamma^S = e^{\Delta S_\mu^m / k_B}$) and the energetic contribution ($\gamma^E = e^{-E_\mu^m / k_B T}$) are also shown for comparison.

**Fig.3** Pressure evolution of the weight fraction ($X_w$) of He, Ne, Ar and Xe in a MORB melt at 2273K. The different quantities contributing to $X_w$ namely, the solubility parameter in the melt ($\gamma_m$), in the noble gas fluid ($\gamma_g$), the ratio ($\gamma_m / \gamma_g$) and the ratio of the densities of the two coexisting phases ($n_g / n_m$) are also shown (for details see text).

**Fig.4** Pressure dependence of the energetic ($\gamma_{m,g}^E$) and entropic ($\gamma_{m,g}^S$) contributions to the solubility parameter ($\gamma_{m,g} = \gamma_{m,g}^E \times \gamma_{m,g}^S$) in a MORB melt (index $m$) at 2273 K and in the coexisting noble gas fluid (index $g$).

**Fig.5** Solubility (in mole fraction) of He, Ne, Ar and Xe in rhyolite, MORB and olivine melts at 2273K as function of pressure. Error bars are evaluated as described in Appendix A.

**Fig.6** Solubility (in weight fraction) of Ar in molten olivine: *triangles* (our evaluation by the TPM at 2273K), *empty circles* (data of Chamorro et al., 1998), and *full circles* (data of Bouhifd and Jephcoat, 2006).

**Fig.7** Solubility (in weight fraction) of Ar in liquid silica: *full triangles* (our evaluation by the TPM at 2600K), *empty triangles* (our evaluation by MD with explicit interface), *empty squares* (data of Chamorro et al., 1996), *full squares* (data of Bouhifd et al., 2008), and *full circles* (low pressure data of Walter et al., 2000). The down arrow indicates approximately the threshold pressure of the Ar solubility drop observed in silica by Chamorro et al. (1996) and Bouhifd et al. (2008).

**Fig.8** Solubility (in weight fraction) of Ar in a basaltic melt: *full triangles* (our evaluation in a MORB melt by the TPM at 2273K), *empty triangles* (our evaluation by MD with explicit interface), *empty circles* (low pressure data of Carroll and Stolper (1993) in olivine tholeiite), and *full circles* (data of White et al. (1989) in olivine tholeiite).

**Fig.9** Noble gas diffusion coefficients in a basaltic melt: *full symbols* with guidelines (our evaluation in a MORB melt at 20 kbar), *error bars* (data of Lux (1987) in a tholeiitic melt), *dotted* line (data of Nowak et al. (2004) for Ar in a synthetic tholeiitic melt), and *dashed* line (diffusion of $CO_2$ in MORB calculated by Guillot and Sator (2011)).

**Fig.10** Noble gas-ion distances in the three simulated melts (T = 2273 K and P~0) as function of the corresponding Lennard-Jones diameters (see text): *circles* (rhyolite), *squares* (MORB) and *triangles* (olivine). In each panel the line expresses the equality $R_{X\text{-ion}} = \sigma_{X\text{-ion}}$.

**Fig.11** Noble gas-oxygen pair distribution functions in rhyolite (*full* curves), MORB (*dashed* curves) and olivine (*dotted* curves) melts at 2273 K and P = 0 and 30 kbar.

**Fig.12** Number of oxygen atoms, $N_O$, in the first solvation shell around a noble gas as function of melt composition and pressure.

**Fig.13** Free volume accessible to a noble gas atom as function of melt composition and pressure: *full* curves (in rhyolite), *dashed* curves (in MORB) and *dotted* curves (in olivine). The case of Ne is not displayed for clarity (very similar to He). The percentage of free volume is defined by $(V-V_{ex})/V$ where V is the volume of melt and $V_{ex}$ is the excluded volume (obtained from the mapping described in section 2.1).

**Fig.A1** Evolution as function of running time of the solubility parameter $\gamma_m$ for He, Ne, Ar and Xe in a MORB melt at 50 kbar and 2273 K

**Fig.A2** Evolution under pressure of the relative error $\Delta\gamma/\gamma$ (where $\Delta\gamma$ is the square root of the variance) for He, Ne, Ar and Xe in a MORB melt at 2273K and for Ar in its parent fluid at 2273K (labeled Ar/Ar in the figure).

**Fig.A3** Noble gas-oxygen and noble gas-silicon pair distribution functions in a MORB melt evaluated by the TPM (*full* curves) and by a MD simulation where a noble gas atom is diluted in the silicate melt (*dotted* curves).

**Fig.B1** Compressibility of the simulated noble gas fluids along the isotherm 2273K using the Tang - Toennies pair potential (2003). For Ar, the results obtained with the pair potential of Ross et al. (1986) is also shown for comparison (see the *empty* triangles). The insert presents a comparison between simulation data for He (*dots*), Ne (*squares*) and Ar (*triangles*) along the isotherm 300 K and the experimental data of the literature (*full* curves: Dewaele et al. (2003) for He and Ne and Tegeler et al. (1999) for Ar).

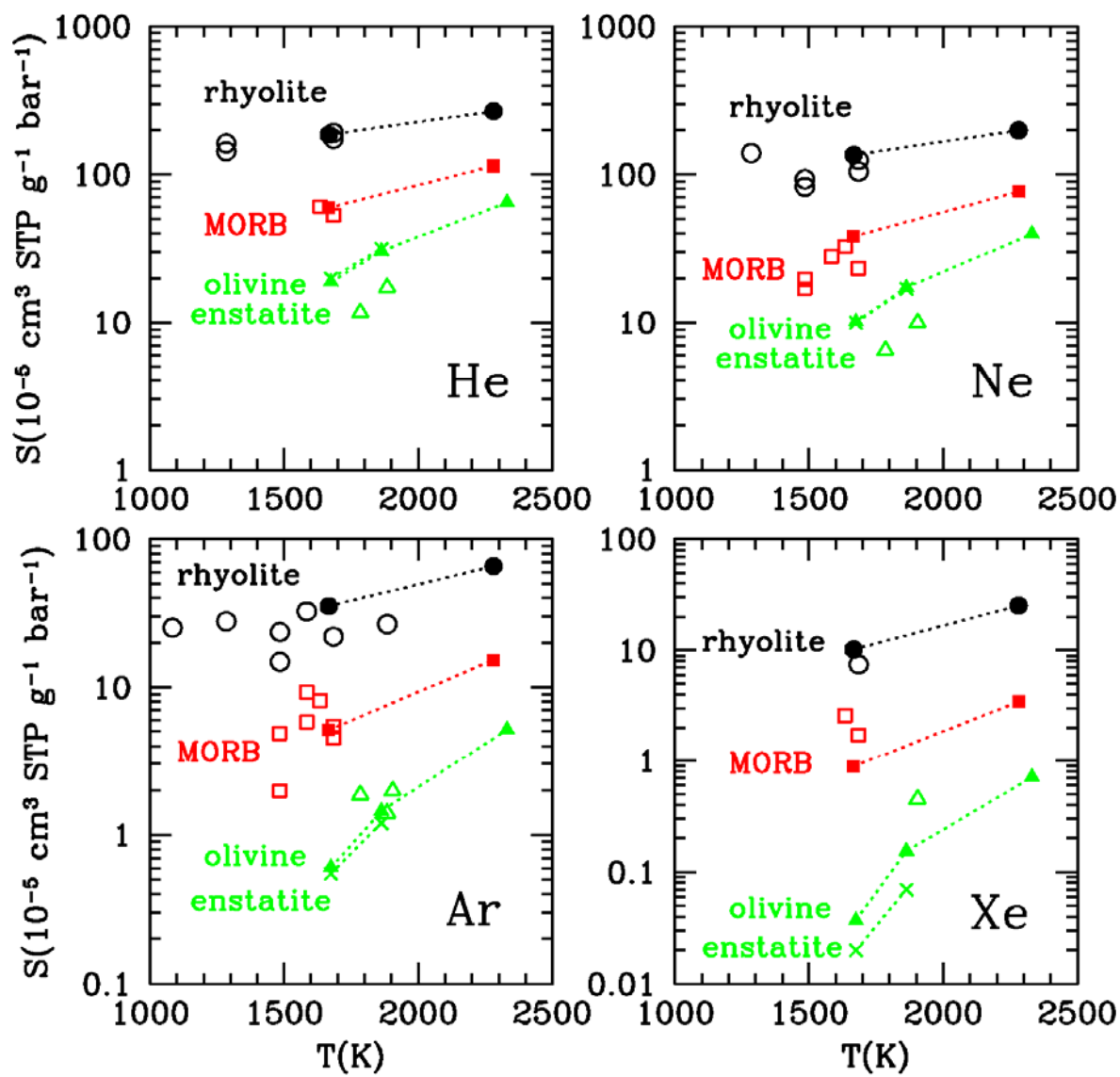

**Fig.1**

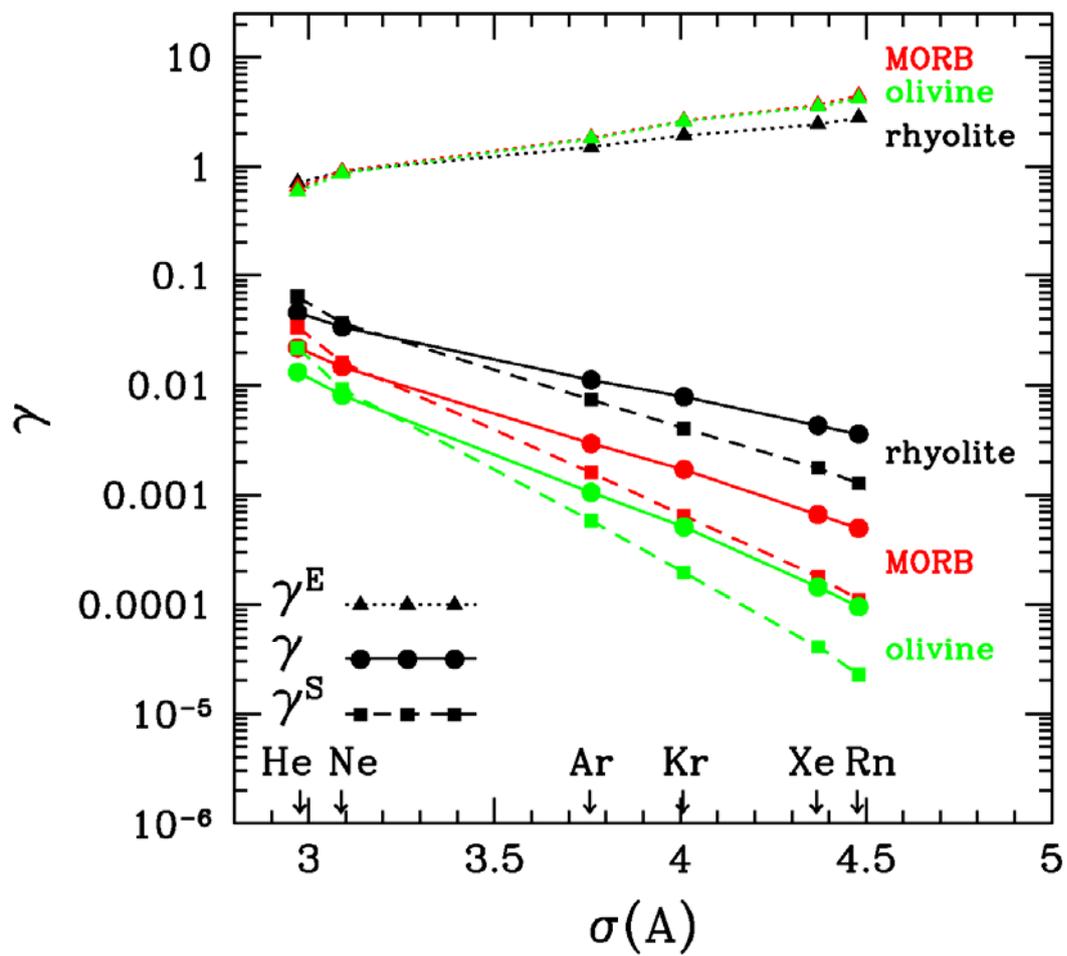

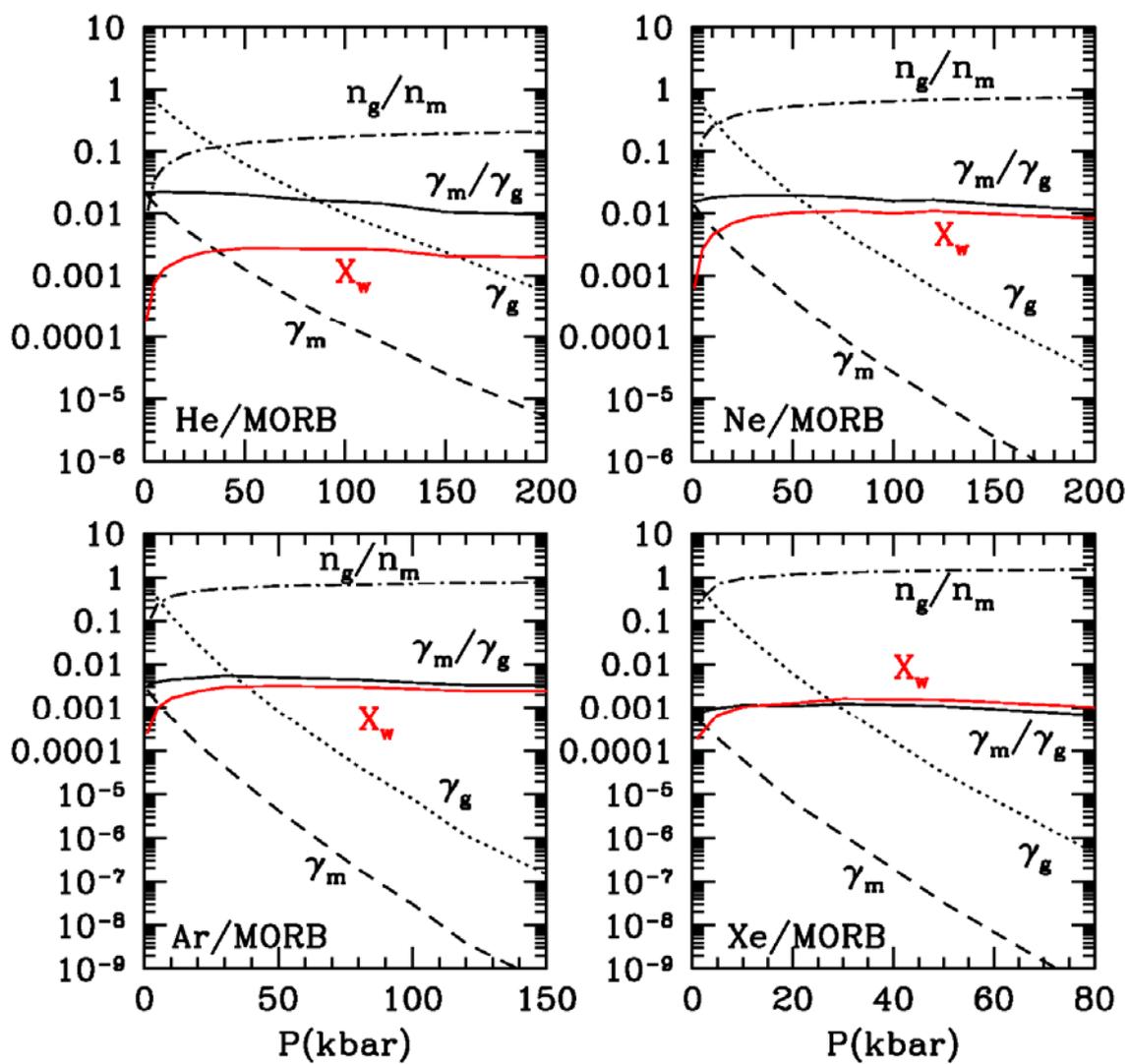

**Fig.3**

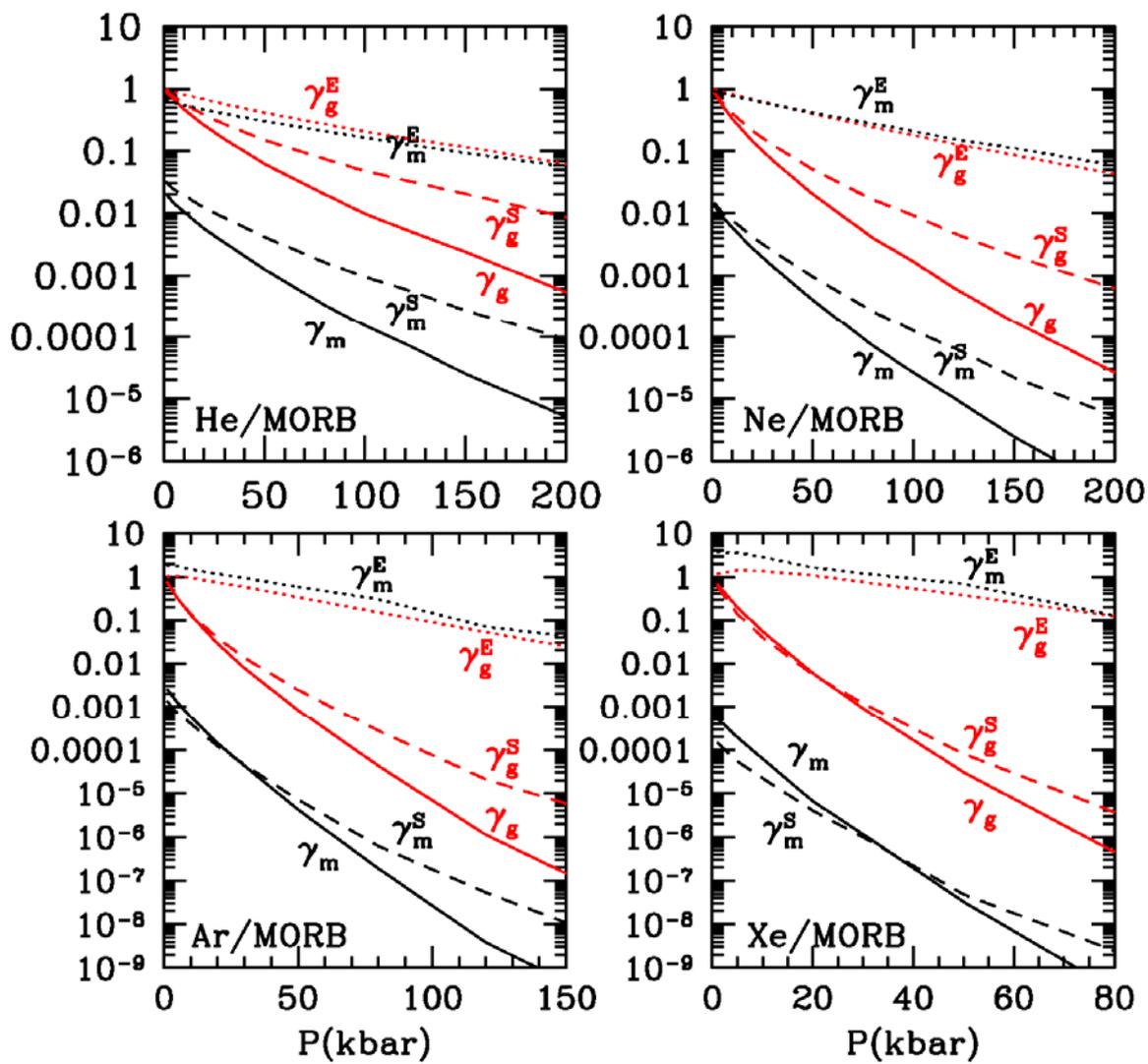

**Fig.4**

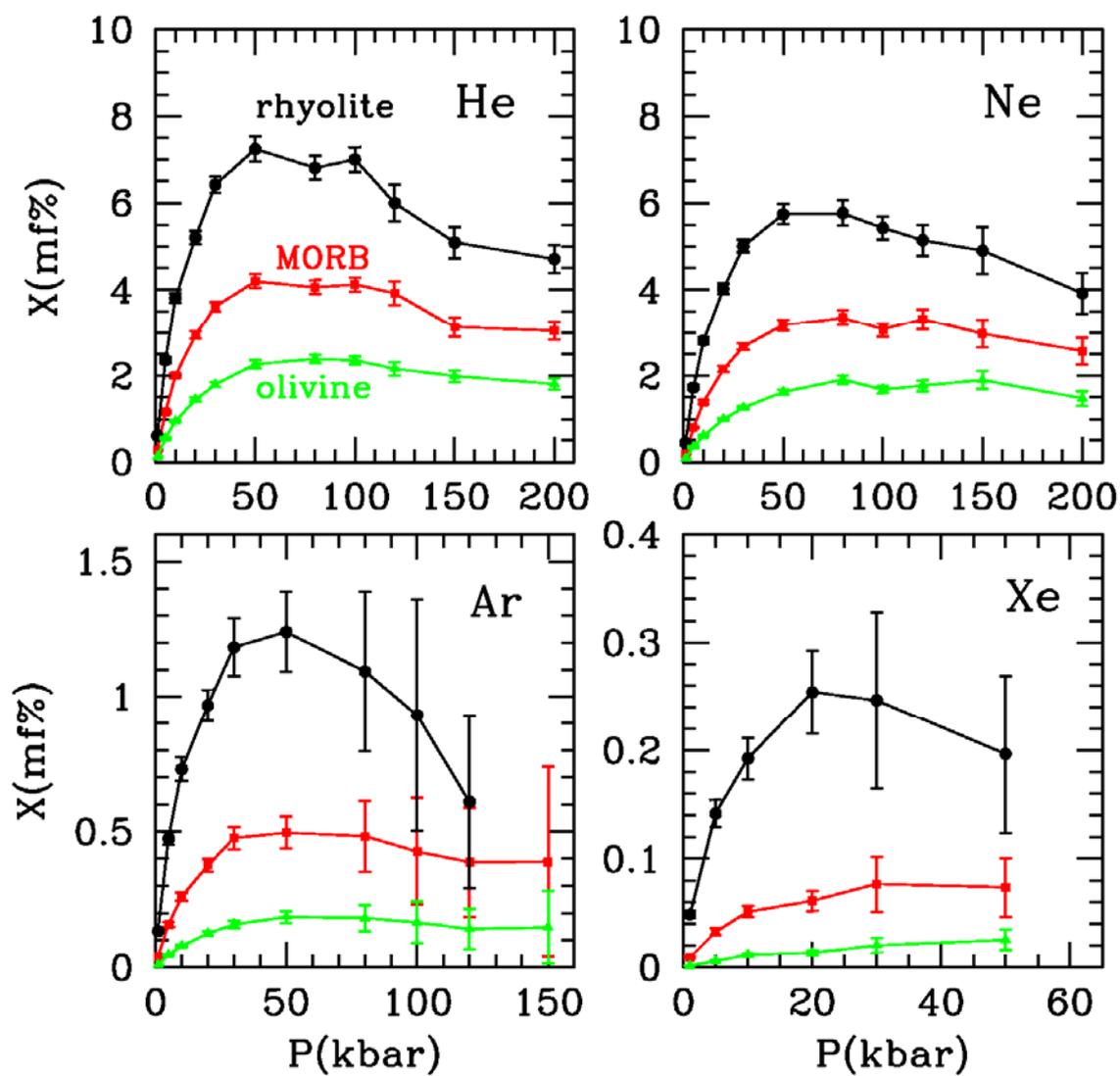

**Fig.5**

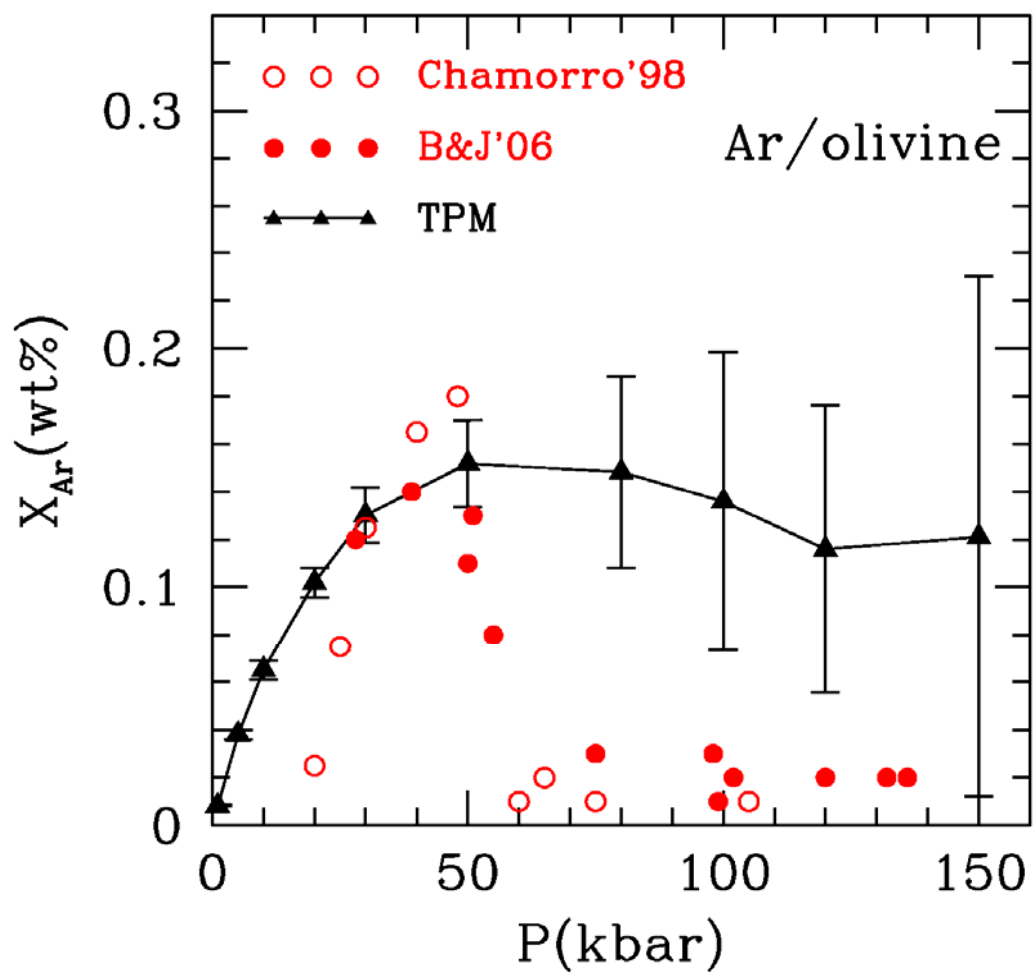

**Fig.6**

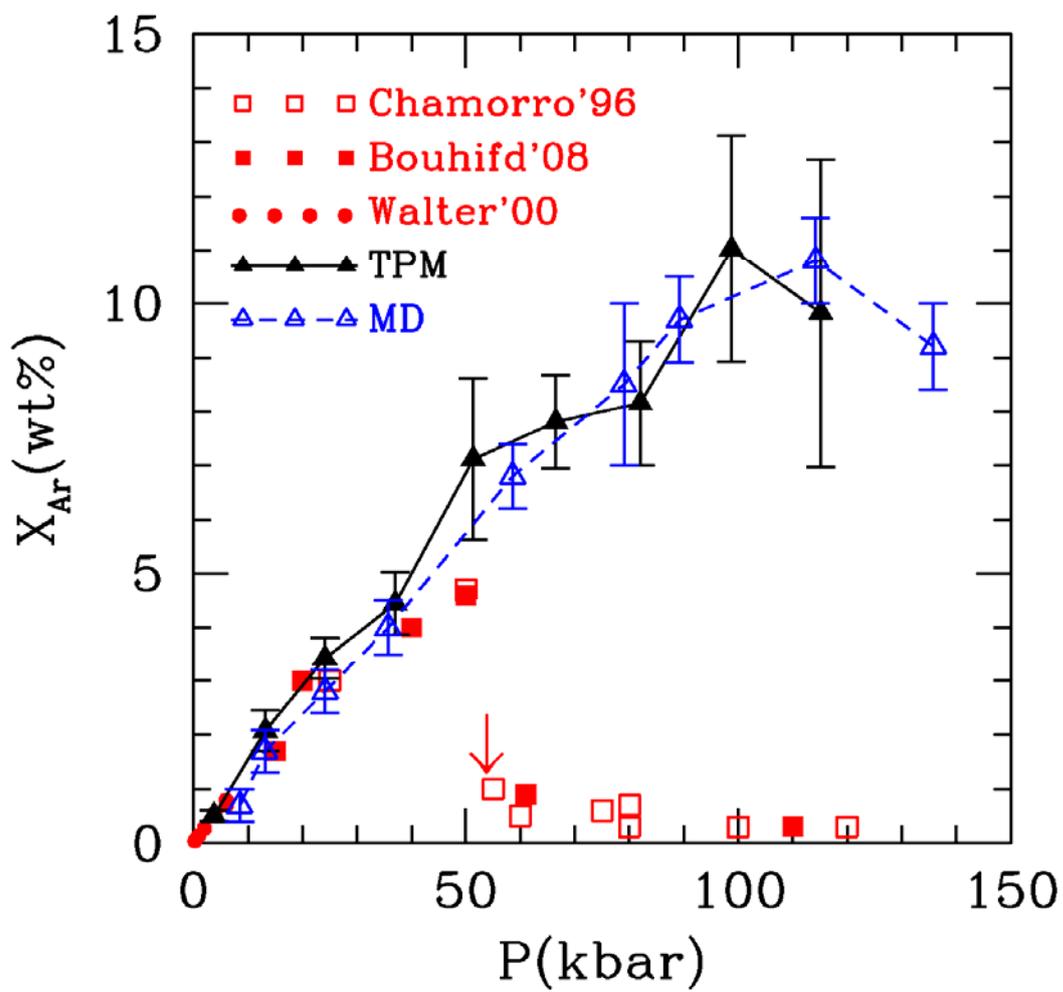

**Fig.7**

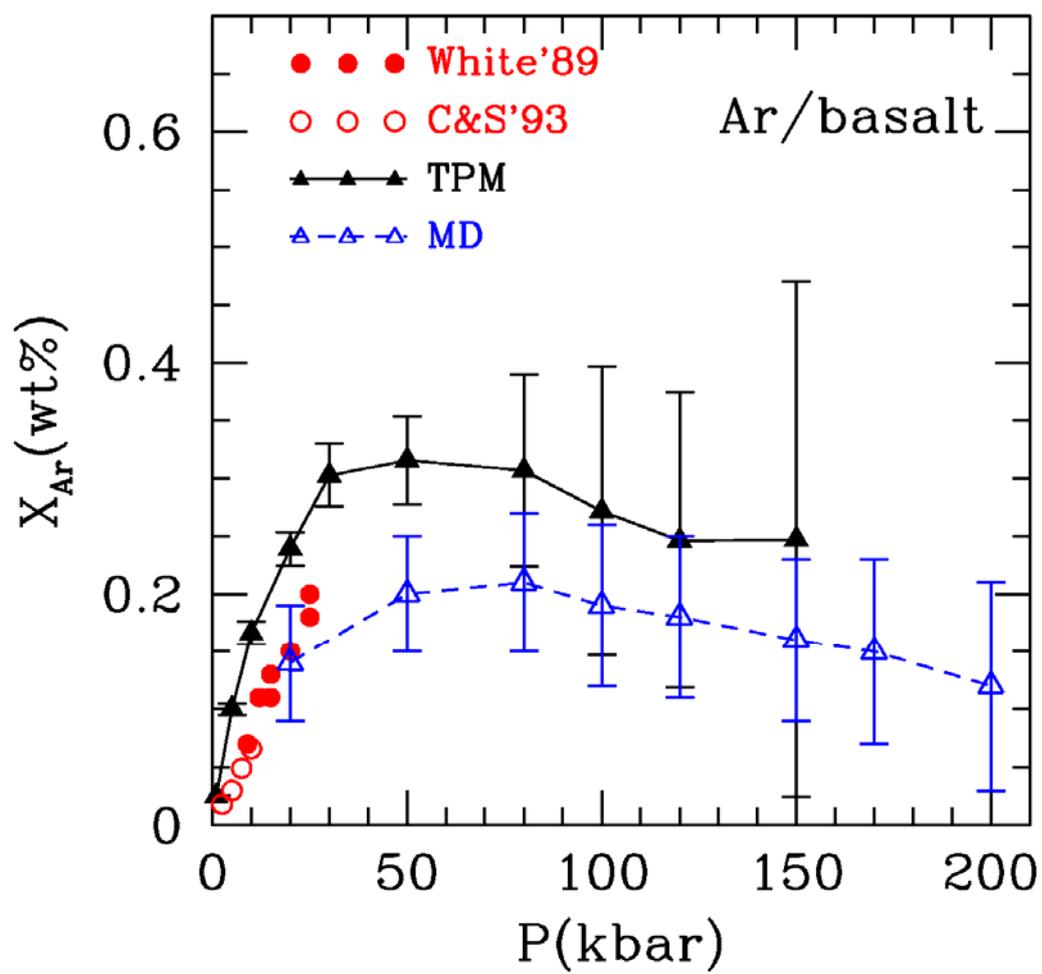

**Fig.8**

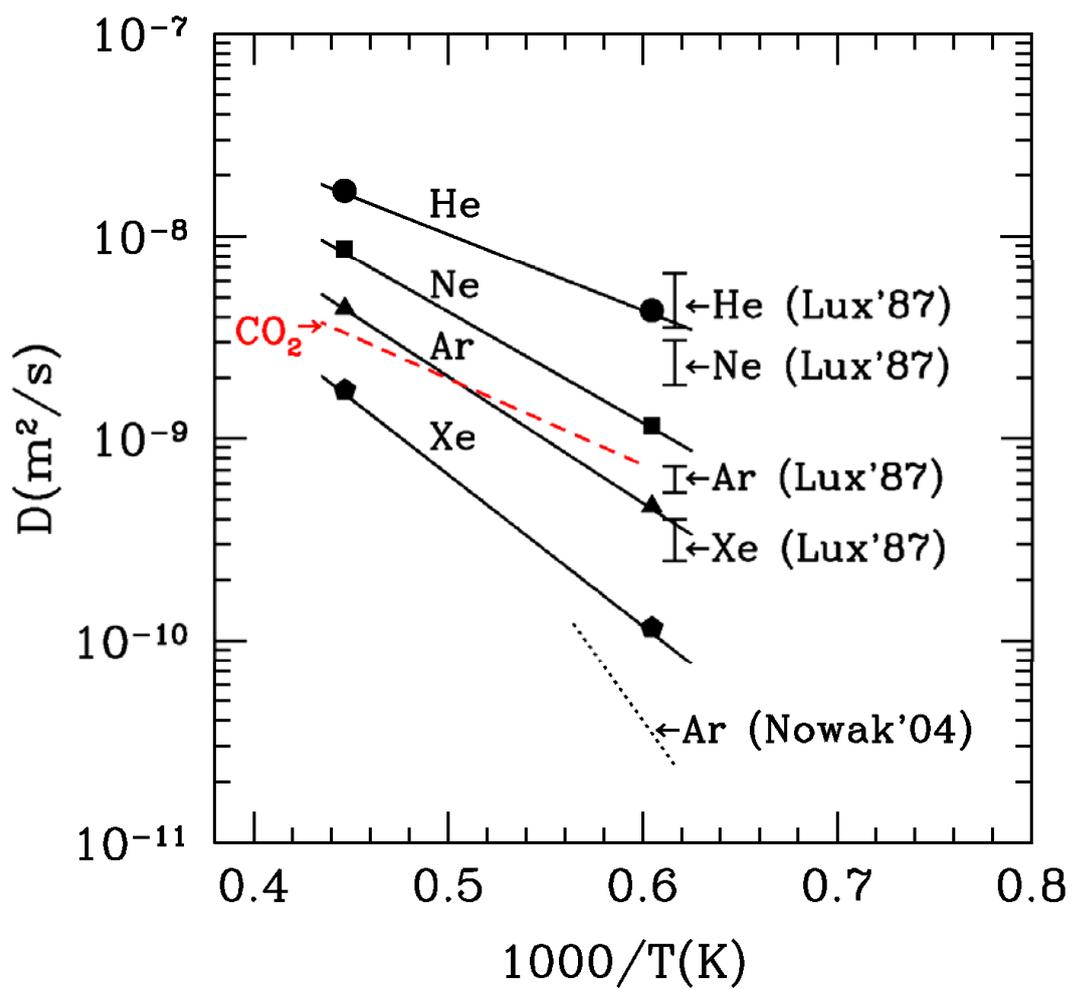

**Fig.9**

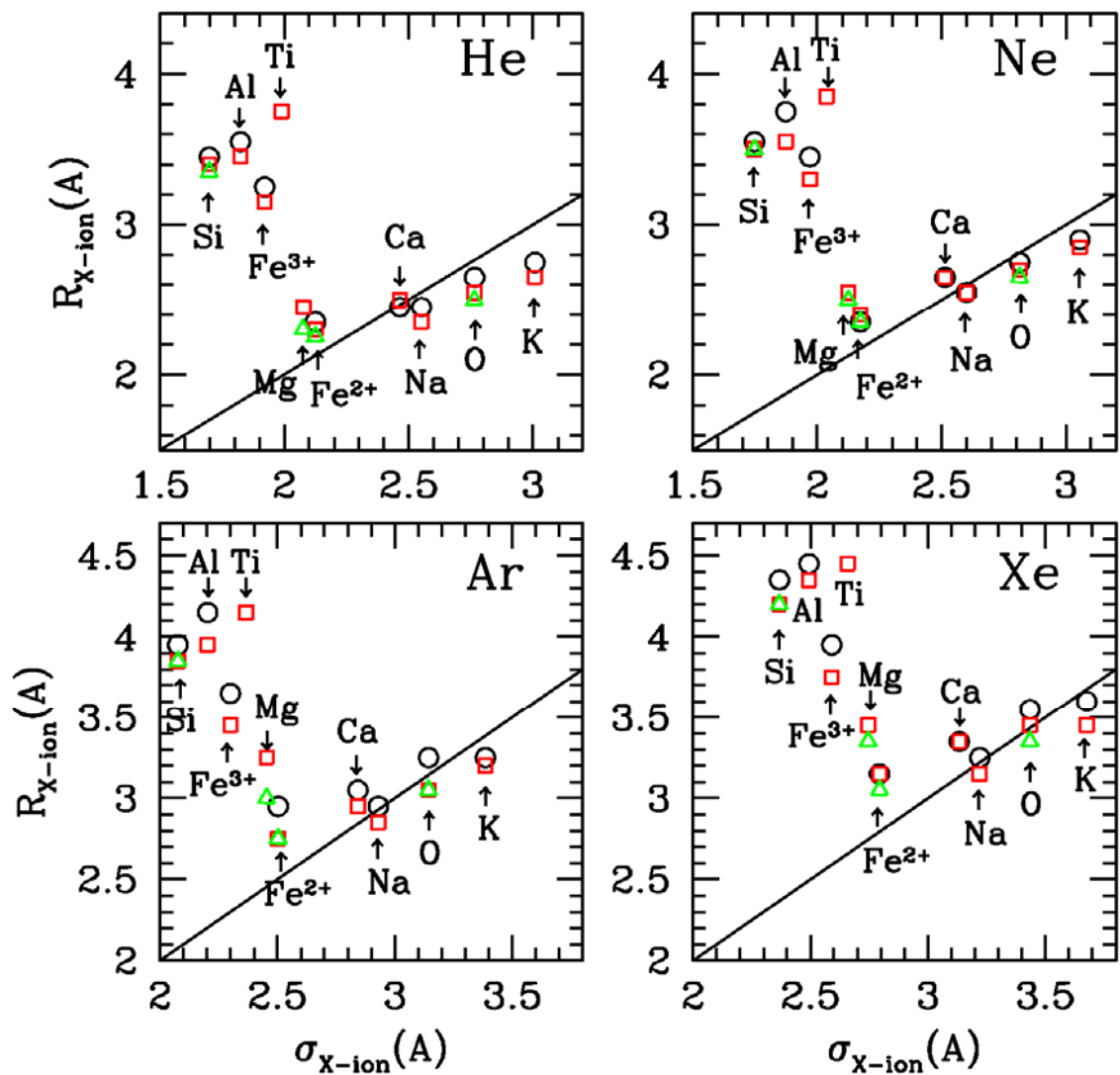

**Fig.10**

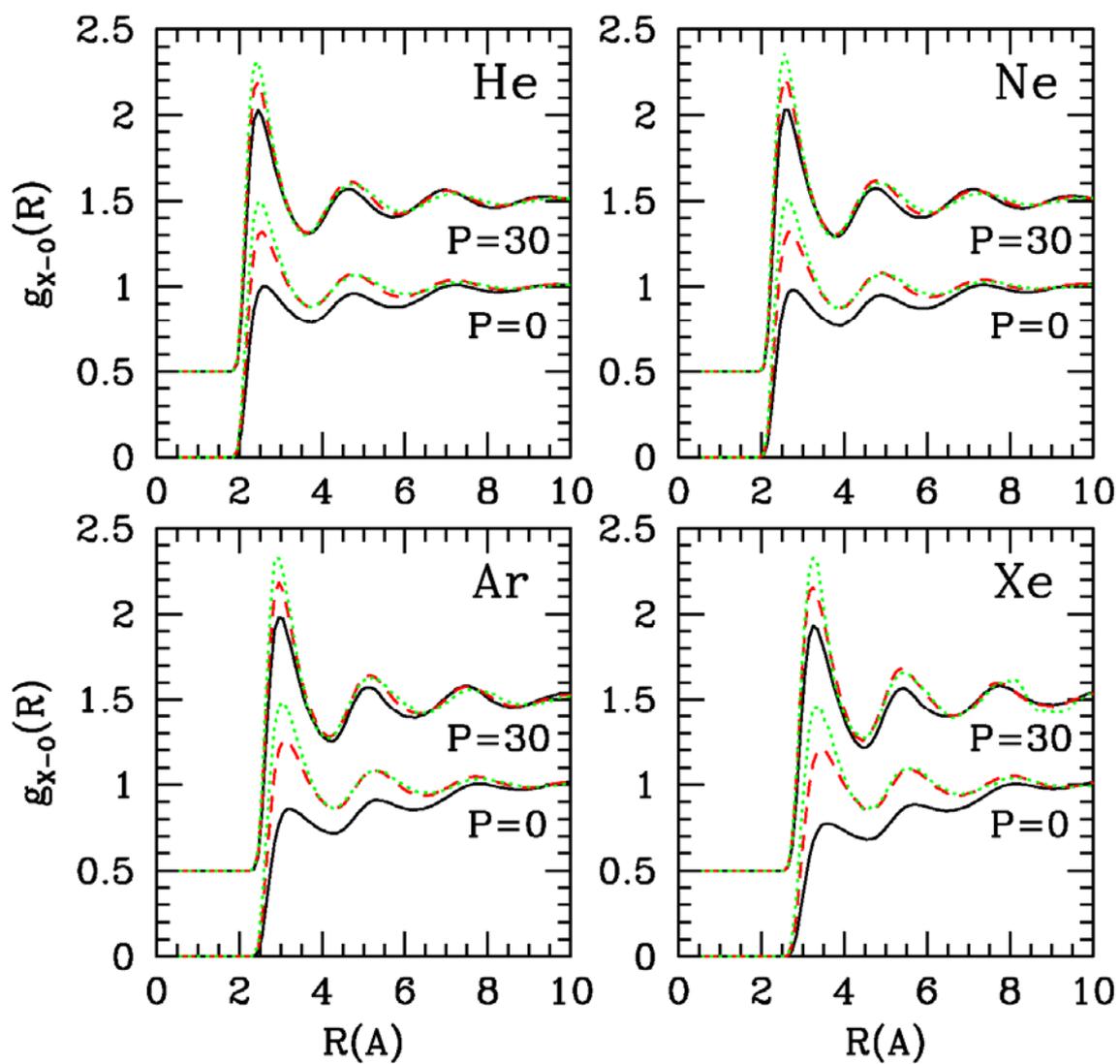

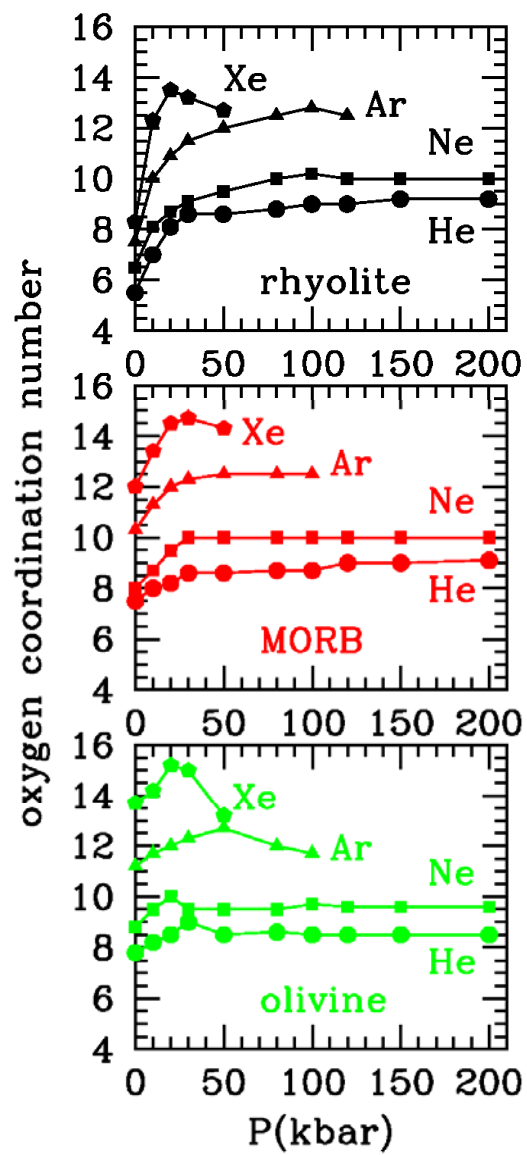

**Fig.12**

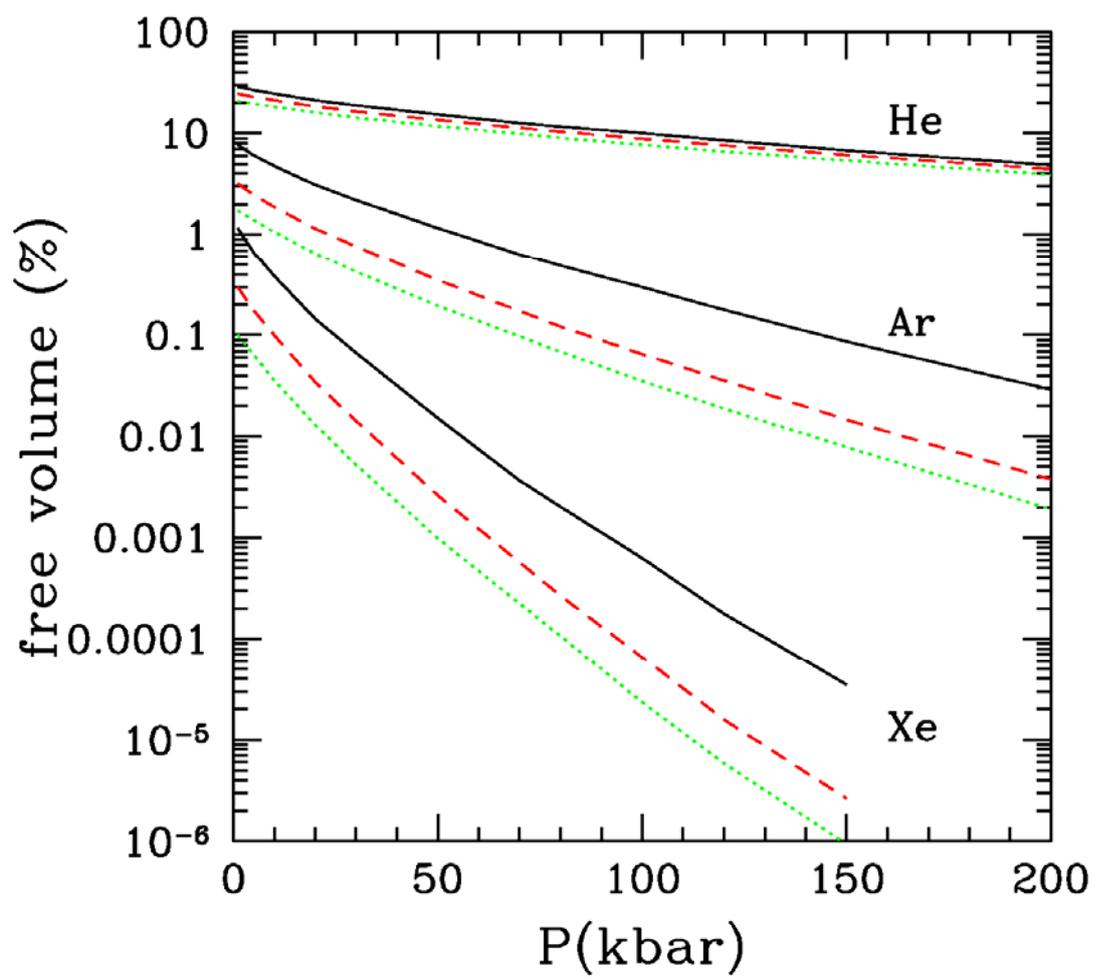

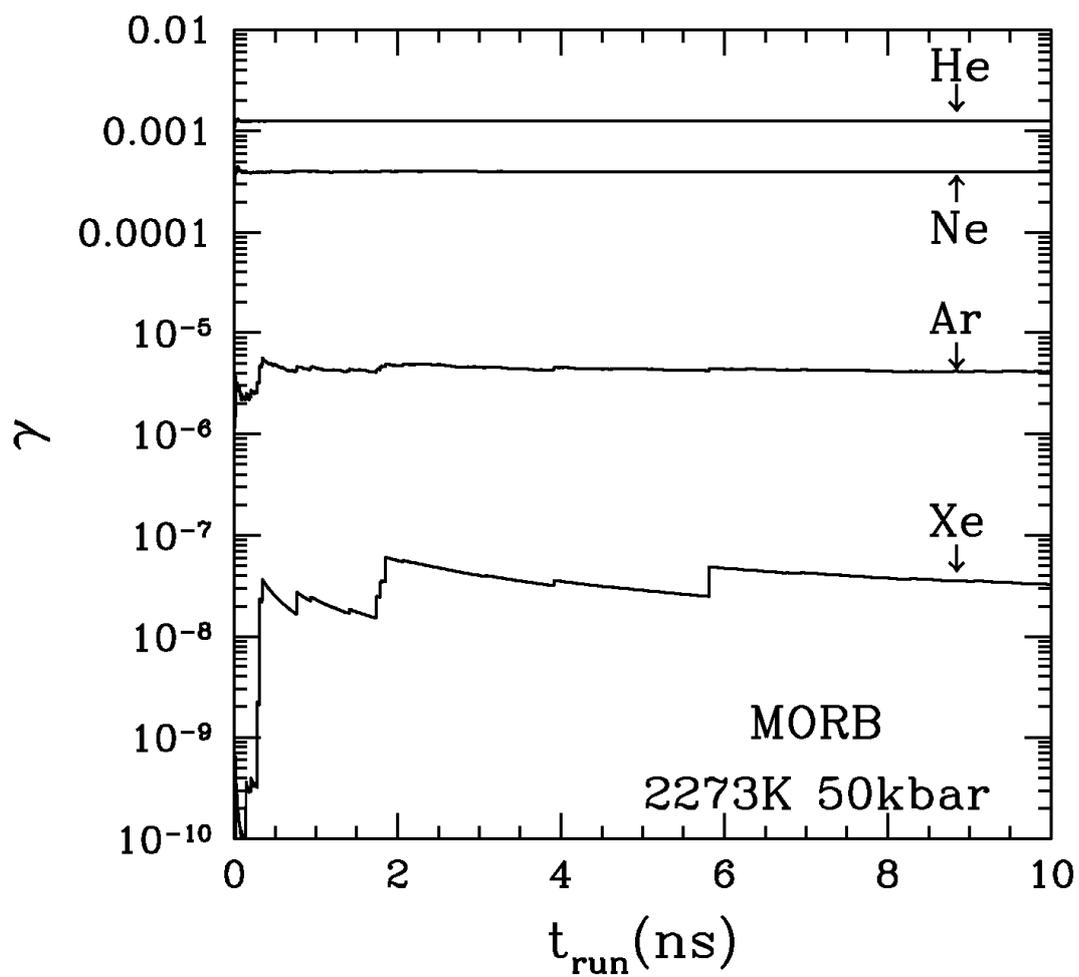

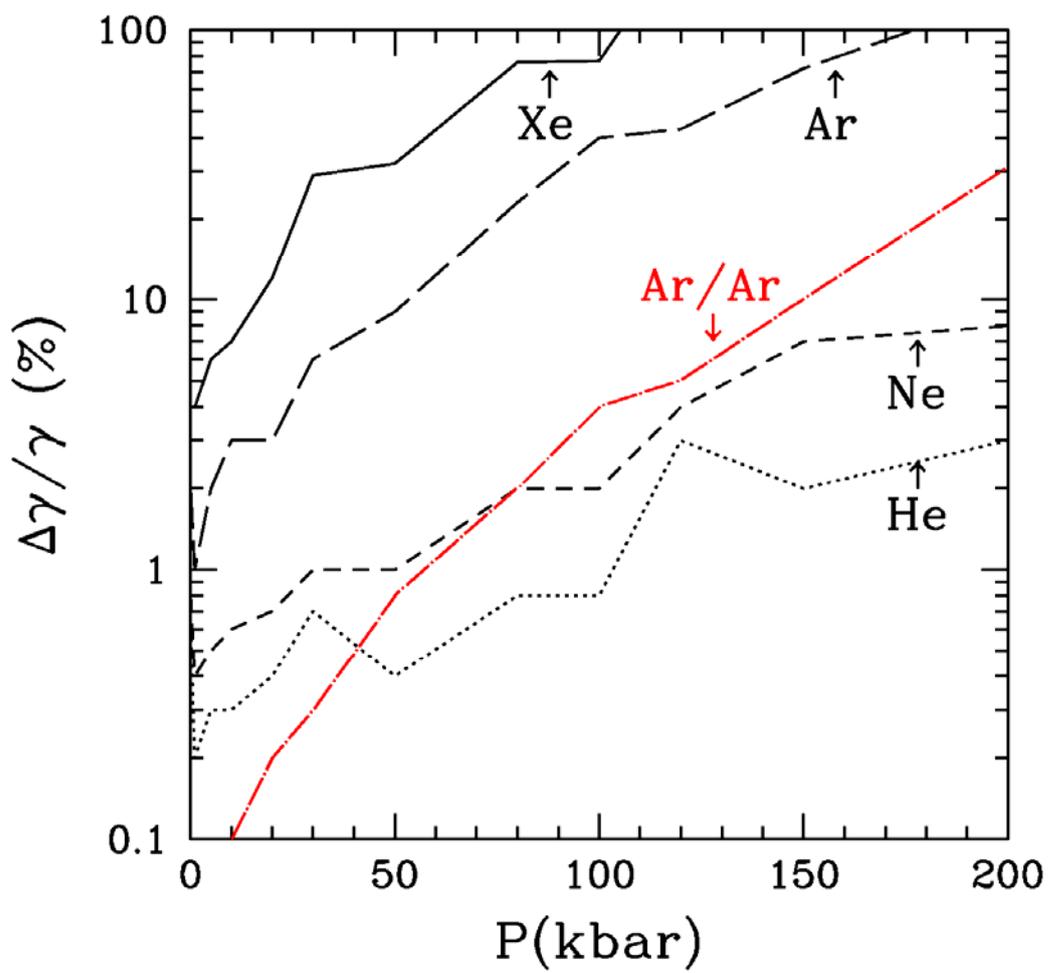

**Fig.A2**

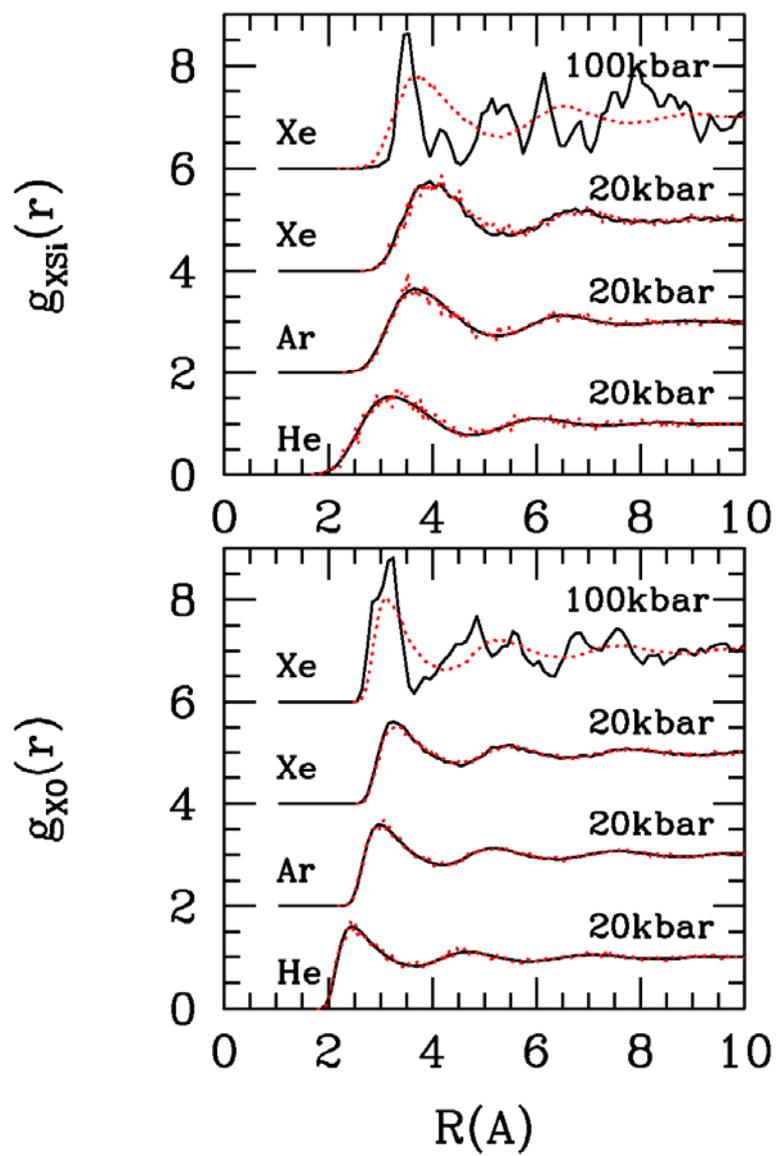

**Fig.A3**

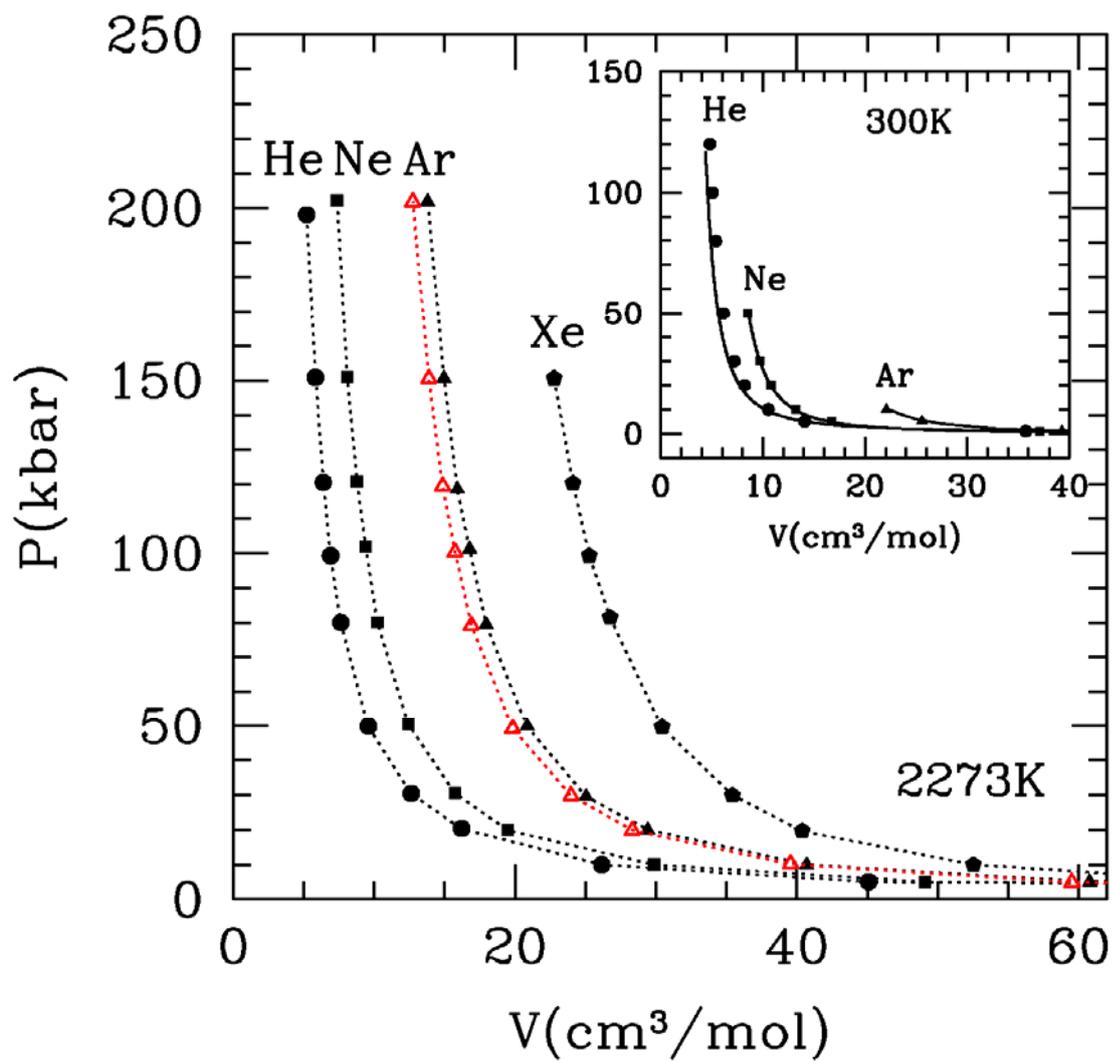

**Fig.B1**